\newif\ifemulate
\newif\ifastroph

\emulatetrue

\astrophtrue

\ifemulate
	\documentclass[iop,numberedappendix,twocolappendix]{emulateapj}
\else
	\documentclass[manuscript]{aastex}
\fi

\newcommand{\myemail}{mulders@lpl.arizona.edu}

\slugcomment{Manuscript in prep. for ApJ}

\shorttitle{Planet radius distribution}
\shortauthors{Mulders et al.}

\usepackage{natbib,graphicx,xspace,amsmath}
\newcommand{\Teff}{\ensuremath{T_{\rm eff}}\xspace}

\newcommand{\SNR}{{\rm S/N}\xspace}

\ifemulate

	\newcommand{\fighalfwidth}{0.495}
\else

	\newcommand{\fighalfwidth}{0.495}
\fi

\begin{document}

\title{An Increase in the Mass of Planetary Systems around Lower-Mass Stars}

\author{Gijs D. Mulders\altaffilmark{1}, Ilaria Pascucci\altaffilmark{1}, and D\'aniel Apai\altaffilmark{1,2}}
\affil{Lunar and Planetary Laboratory, The University of Arizona, Tucson, AZ 85721, USA}
\email{\myemail}
\altaffiltext{1}{Earths in Other Solar Systems Team, NASA Nexus for Exoplanet System Science}
\altaffiltext{2}{Department of Astronomy, The University of Arizona, Tucson, AZ 85721, USA}

\begin{abstract}
Trends in the planet population with host star mass provide an avenue to constrain planet formation theories. We derive the planet radius distribution function for Kepler stars of different spectral types, sampling a range in host star masses. We find that M dwarf stars have 3.5 times more small planets ($1.0-2.8~R_\oplus$) than main-sequence FGK stars, but two times fewer Neptune-sized and larger ($>2.8 R_\oplus$) planets. We find no systematic trend in the planet size distribution between spectral types F, G, and K to explain the increasing occurrence rates.
Taking into account the mass-radius relationship and heavy-element mass of observed exoplanets, and assuming those are independent of spectral type, we derive the inventory of the heavy-element mass locked up in exoplanets at short orbits. The overall higher planet occurrence rates around M stars are not consistent with the redistribution of the same mass into more, smaller planets.
At the orbital periods and planet radii where Kepler observations are complete for all spectral types, the average heavy-element mass locked up in exoplanets increases roughly inversely with stellar mass from $4 ~M_\oplus$ in F stars to $5 ~M_\oplus$ in G and K stars to $7 ~M_\oplus$ in M stars. 
This trend stands in stark contrast with observed protoplanetary disk masses that decrease towards lower mass stars, and provides a challenge for current planet formation models. Neither models of in situ formation nor migration of fully-formed planets are consistent with these results. Instead, these results are indicative of large-scale inward migration of planetary building blocks --- either through type-I migration or radial drift of dust grains --- that is more efficient for lower mass stars, but does not result in significantly larger or smaller planets.
\end{abstract}


\section{Introduction}
The \textit{Kepler} spacecraft has unearthed a population of planets at short orbital periods that orbit the majority of stars in our galaxy \cite[e.g.][]{Borucki:2011cp,2012ApJS..201...15H,2013ApJ...766...81F,2015ApJ...809....8B}. This population is significantly distinct in planet size, orbital period, and bulk composition from previously known planetary systems, including our solar system. The formation history of these ``Super-Earths'' or ``mini-Neptunes'' is still open to debate. They may have formed in situ in an up-scaled version of our solar system \citep{2013MNRAS.431.3444C,2013ApJ...775...53H}; they may have formed at larger distances from the star and migrated inward \citep[e.g.][]{2013ApJ...764..105S}; or they formed through a combination of both processes \citep[e.g.][]{2012ApJ...751..158H,2013A&A...558A.109A, 2014A&A...569A..56C}.

Variations in the planet population with host star properties provide an avenue of discriminating between planet formation theories. Radial velocity surveys indicate that the occurrence of giant planets correlates strongly with host star mass \citep{Johnson:2010gu} and metallicity \citep[e.g.][]{2000A&A...363..228S}. These results are consistent with the core-accretion paradigm of planet formation, in which larger solid mass in higher-mass and higher-metallicity disks is necessary to form cores massive enough to start run-away gas accretion \cite[e.g.][]{Ida:2004jo,Ida:2005bm,Dawson:2015ha}.

The presence of a large population of planets smaller than Neptune (``failed cores'') at shorter orbital periods does not match the predictions of the core-accretion model \citep{2010Sci...330..653H,2012ApJS..201...15H}.
Their presence does not correlate with stellar metallicity \citep{2008A&A...487..373S, 2012Natur.486..375B, Schlaufman:2015di,Buchhave:2015cg} , though weaker trends with metallicty are evident \citep{Dawson:2015ha,Adibekyan:2015vb}. Surprisingly, the occurrence rate of these sub-Neptune sized planets \textit{increases} for lower-mass stars \citep{2012ApJS..201...15H,Mulders:2015ja}, opposite to the trend for giant planets. Qualitatively, this does match the prediction of in-situ formation models, where more numerous, but smaller planets form in low-mass disks around low-mass stars \citep[e.g.][]{Wetherill:1996ga,Kokubo:2002dz,2007ApJ...669..606R,Ciesla:2015ha}. In agreement with these predictions, a lack of planets larger than $\sim 2.5 ~R_\oplus$ has indeed been noted around the lowest-mass stars in the Kepler sample \citep{2013ApJ...767...95D,2015ApJ...807...45D,2014ApJ...791...10M}. 

In this paper, we place these results into context by investigating how the planet radius distribution varies as a function of spectral type, and hence stellar mass. We explore and reject the hypothesis that higher occurrence rates toward lower mass stars are due to a redistribution of mass into more numerous, smaller planets. We make a quantitative comparison between the close-in planet population and different planet formation models.

\begin{table*}
	\title{Planet Occurrence Rates per KOI}
	\centering
   \begin{tabular}{l l r l | c c l}
   \hline\hline
   KOI & $R_P$ [cm] & $P$ [day] & $f_{\rm occ}$ & \Teff [$K$] & Spectral Type & $f_{\rm occ}$ \\
   \hline
   K00001.01 & 7.875e+09 & 2.471 & 3.989e-05 & 5850 & G & 1.112e-04 \\
   K00002.01 & 1.133e+10 & 2.205 & 3.537e-05 & 6350 & F & 7.988e-05 \\
   K00003.01 & 3.508e+09 & 4.888 & 6.683e-05 & 4777 & K & 4.311e-04 \\
   K00005.01 & 3.801e+09 & 4.780 & 6.554e-05 & 5753 & G & 1.831e-04 \\
   K00007.01 & 2.878e+09 & 3.214 & 5.104e-05 & 5781 & G & 1.426e-04 \\
   ... & ... & ... & ... & ...  & ... & ... \\
   K06239.01 & 1.200e+09 & 406.496 & 4.134e-03 & 5847 & G & 1.058e-02 \\
   K06242.03 & 8.919e+08 & 78.867 & 1.309e-03 & 6907 & F & 3.148e-03 \\
   K06245.02 & 1.184e+09 & 112.317 & 9.466e-04 & 6528 & F & 2.255e-03 \\
   K06246.01 & 1.011e+09 & 9.133 & 1.254e-04 & 6122 & F & 2.861e-04 \\
   K06248.01 & 4.189e+10 & 1.759 & 2.201e-05 & 5721 & G & 6.037e-05 \\
   \hline\hline\end{tabular}
	\par 
	\caption{\ifemulate Planet Occurrence Rates per KOI. \fi 
	The rightmost column lists the occurrence rates calculated for each spectral-type subsample. In case the effective temperature is outside the range for FGKM stars, its spectral type is listed as X and its occurrence as not-a-number (nan). Table \ref{tab:occ} is published in its entirety in the electronic edition of the Astrophysical Journal. A portion is shown here for guidance regarding its form and content.}
	\label{tab:occ}
\end{table*} 	

\section{Planet occurrence rates}
Planet occurrence rates are calculated using the methodology described in \cite{Mulders:2015ja}. 
We use the planet candidate sample from \cite{Mullally:2015iq}, which is currently the preferred catalogue for occurrence rates studies \citep[e.g.][]{2015ApJ...809....8B}.
In short, for every main-sequence star observed by the \textit{Kepler} spacecraft, we calculate whether a planet of given radius, $R_P$, and orbital period, $P$, could be detected based on the combined signal to noise of all transits during the time the star was observed. The stellar parameters (\Teff, $R_\star$, $M_\star$) are taken from \cite{Huber:2014dh}. The stellar noise during the duration of a single transit is interpolated from the Combined Differential Photometric Precision \citep{Christiansen:2012bz} at 3, 6, and 12 hour time scales, downloaded from the MAST archive\footnote{http://archive.stsci.edu/kepler/data\_search/search.php} on December 12th 2013. The occurrence rate $f_{\rm occ}$ in the range $\{R_P, P, \Teff\}$ is the ratio of the number of KOIs over the number of stars were planets are detectable, multiplied by the geometric transit probability. We treat all KOIs as real planets, consistent with the low false positive rate in the Kepler data \citep{2013ApJ...766...81F}. In the following paragraph, we describe in detail where we deviate from the assumptions in \cite{Mulders:2015ja}.

We use the KOI list spanning quarters Q1-Q16 (4 years) from \cite{Mullally:2015iq}, replacing the Q1-Q8 list (2 years) from \cite{2014ApJS..210...19B} in our previous work. Aside from doubling the observing time baseline, this new version of the pipeline has an improved detection efficiency at low transit signal-to-noise (S/N), adding many smaller planet candidates. We use the empirical detection efficiency from planet injection test by \cite{Christiansen:2015ge}. This efficiency is described by a cumulative gamma function\footnote{Offset by a \SNR of 4.1: $p=F(\SNR-4.1|a,b)$} with $a=4.35$ and $b=1.05$, that replaces the linear ramp between S/N of 6 and 12 in our previous work. We compared the tabulated signal-to-noise and transit time for all KOIs to the calculated one, and we no longer find a systematic offset. This eliminates the need for the free scaling parameter described in appendix B.3 of \cite{Mulders:2015ja}. We also assume a transit duration based on an uniform impact parameter distribution, rather than a zero impact parameter. The sample is subdivided by spectral type according to effective temperature, \Teff, from Table 5 in \cite{Pecaut:2013ej}, where we use temperatures of $3865 ~K$, $5310 ~K$, $5980 ~K$, and $7320 ~K$ as upper bounds for M, K, G , and F stars, respectively. 
Occurrence rates for each KOI in both the full sample and spectral type subsamples are provided in Table \ref{tab:occ}.

\begin{figure}
	\includegraphics[width=\linewidth]{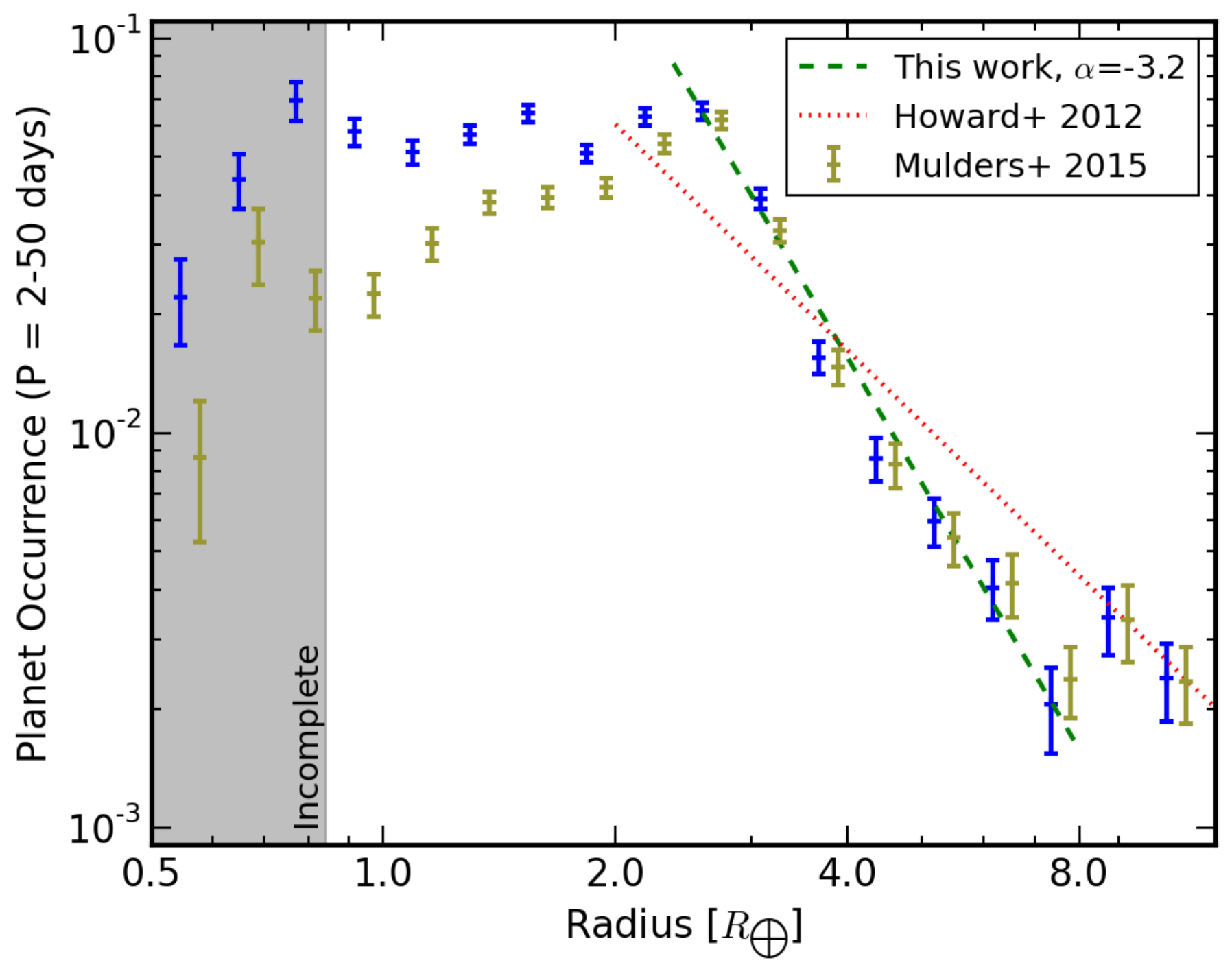}
	\caption{Planet radius distribution for orbital periods between 2 and 50 days for all main-sequence stars in the Kepler sample. $1-\sigma$ error bars are calculated from the square root of the number of KOIs in each bin. The sample of stars becomes incomplete below $\sim 1 ~R_\oplus$. The red dotted line shows the best-fit power-law in planet radius ($k_R=2.9, \alpha=-1.9$) derived by \cite{2012ApJS..201...15H}. The green dashed line shows the best-fit power-law between $2.4-8 ~R_\oplus$ with $k_R=20$, $\alpha=-3.3$. The occurrence rates from \cite{Mulders:2015ja}, based on the first eight quarters and shifted by 0.06 dex in planet radius for clarity, are shown for reference. These rates show good agreement for planets larger than $2 ~R_\oplus$, while our previous work underestimated the amount of smaller planets.
	\label{f:radius}
	}
\end{figure}

\begin{figure}
	\includegraphics[width=\linewidth]{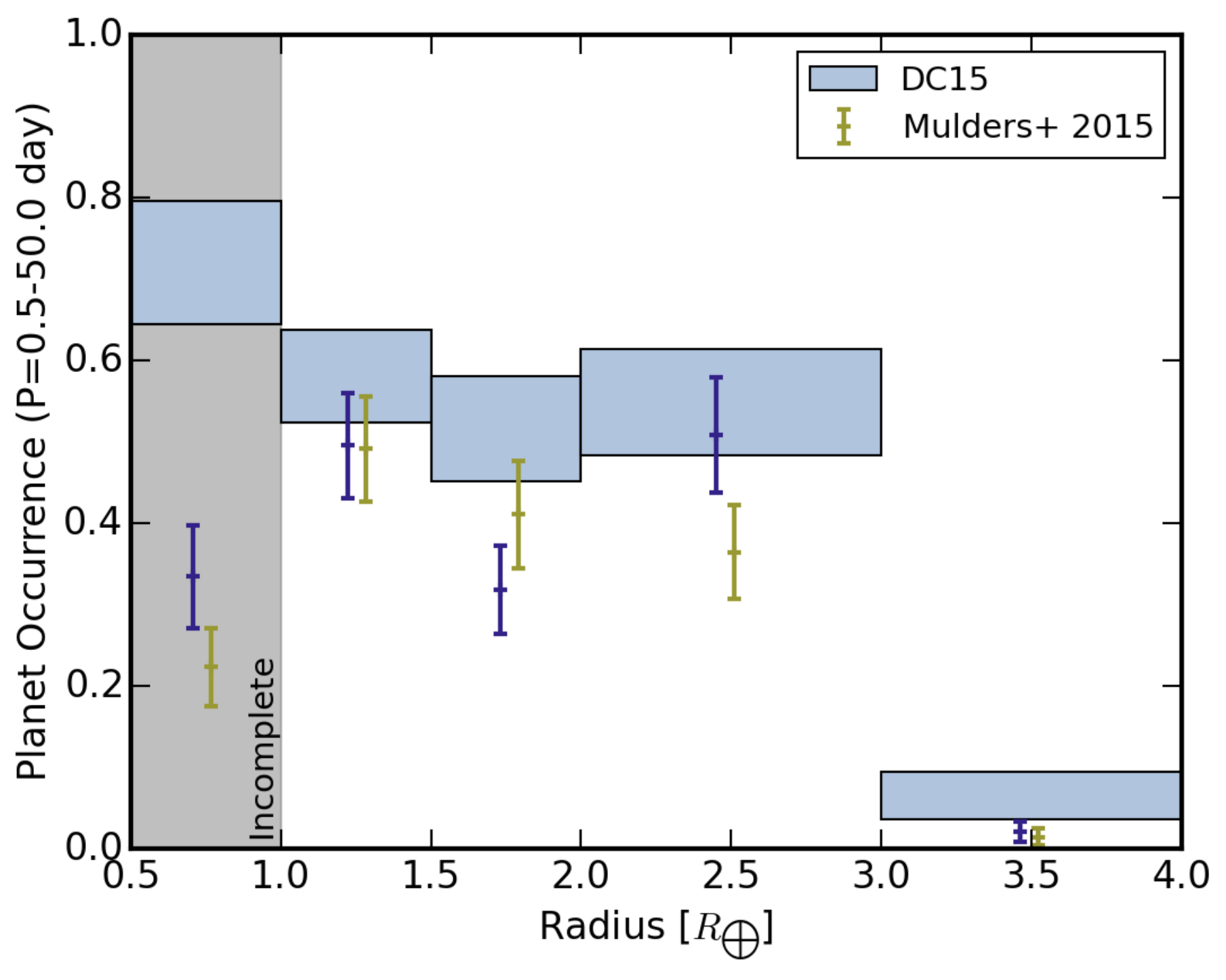}
	\caption{Planet radius distribution for orbital periods up to 50 days for the cool star sample, compared to \cite{Mulders:2015ja} (yellow symbols, offset by $0.06 ~R_\oplus$ for clarity) and \cite{2015ApJ...807...45D} (the height of the blue box represents the 1-$\sigma$ error). The improved estimate of the detection efficiency in \cite{2015ApJ...807...45D} yields higher planet occurrence rates for the coolests stars.
	1-$\sigma$ error bars are calculated from the square root of the number of KOIs in each bin. The sample of stars becomes incomplete below $\sim 1 ~R_\oplus$.
	\label{f:coolstars}
	}
\end{figure}

\subsection{Planet radius distribution}
We calculate the planet radius distribution between orbital periods of 2 to 50 days. The inner edge of 2 days is chosen because there is a clear lack of large ($>2 ~R_\oplus$) planets at shorter orbits, which might be a signature of photo-evaporation \citep[e.g.][]{SanchisOjeda:2014gi}. To include planets down to one Earth radius even around M stars, we choose an outer period cut of 50 days.
We use orbital period, rather than semi-major axis, to define this region because \cite{Mulders:2015ja} have shown that the characteristic feature in the occurrence rates with distance from the star scales with orbital period when considering different stellar-mass sub-samples. 
This range encompasses the peak in detected KOIs, which lies roughly at 10 days, optimizing the detection statistics.

We construct the planet radius distribution by summing the occurrence rates in an orbital period and planet radius bin. The default orbital period bin is 2-50 days as described above, and planet radius bins are equally spaced in log space with 4 bins per factor 2 increment. Errors are given as the square root of number of KOIs in each bin. Figure \ref{f:radius} shows the planet radius distribution for the entire Kepler sample of main-sequence stars.
These rates show overall good agreement with those in \cite{2012ApJS..201...15H}, who fit a power-law in planet radius for planets within $P<50$ days between radii $2-22.6 ~R_\oplus$ with an index $\alpha=-1.9$. Our finer binning, larger number of KOIs, and completeness down to smaller planets reveals additional structure in this distribution: a plateau below $2.8 ~R_\oplus$ \citep[see also][]{2013ApJ...770...69P}, a steep dropoff towards large planet radii, and a flattening between $8-16 ~R_\oplus$. We identify a steeper slope of $\alpha=-3.2$ in a narrower range of radii, $2.4-8 ~R_\oplus$, compared to \cite{2012ApJS..201...15H}.

The occurrence rates are consistent with our previous work \citep{Mulders:2015ja} for planets larger than $2 ~R_\oplus$, but significantly higher for smaller planets. The addition of many smaller planet candidates in the Q1-Q16 catalogue boosts occurrence rates in this regime, and our previous estimate of the detection efficiency (a linear ramp between SNR of 6 to 12) was not sufficient to correct for this. We also compare occurrence rates for the cool star sub-sample ($\Teff<4000 K$) to those of \cite{2015ApJ...807...45D} and \cite{Mulders:2015ja} in Figure \ref{f:coolstars}. The rates are consistent with our previous work within errors. The rates from \cite{2015ApJ...807...45D} are systematicaly higher, most likely due to their better estimate of the detection efficiency in their own pipeline, which does not reach 100\%, boosting occurrence rates for planets at all signal-to-noise ratios. \cite{Christiansen:2015ge} found that the detection efficiency is different for non-FGK stars, but the detection efficiency for M stars could not be accurately determined. This lower detection efficiency may indicate that the occurrence rates for M stars are currently underpredicted.

\begin{figure}
	\includegraphics[width=\linewidth]{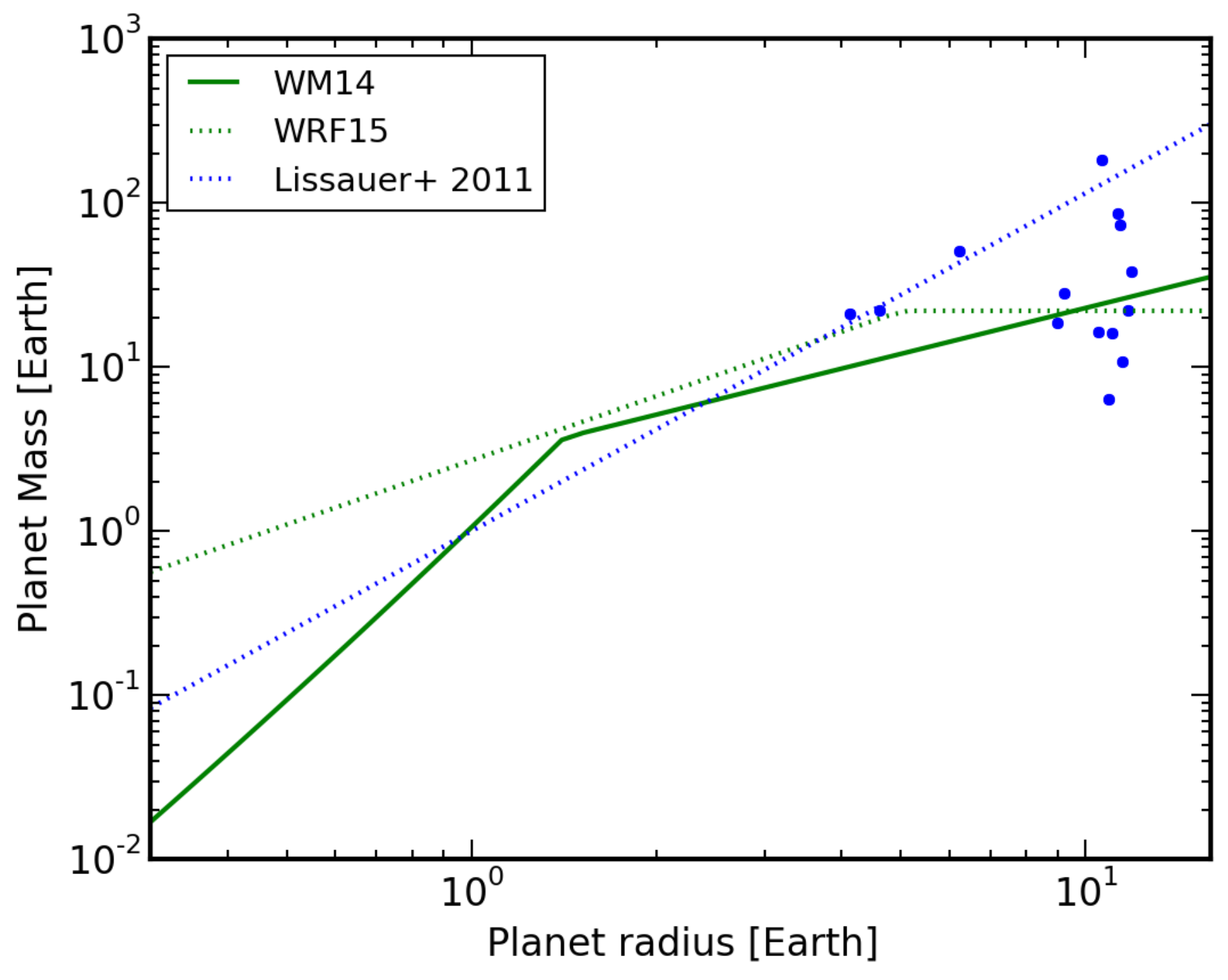}
	\caption{Planet mass-radius relationships used in this work: \cite{2014ApJ...783L...6W} (WM14) and \cite{2015arXiv150407557W} (WRF15). The blue points are the heavy element masses from \cite{Miller:2011bq}. The WM14 relation matches well with the median core mass of giant planets. The relation from WRF15 has a steeper slope and predicts higher planet masses overall, so we use a maximum mass of 22 $M_\oplus$ to provide a better estimate of
 the heavy-element mass function for giant planets. The mass-radius relation from \cite{Lissauer:2011fj} based on the Solar System is shown for comparison.
	\label{f:MR}
	}
\end{figure}

\subsection{Conversion to planet heavy-element mass}
To test if the higher occurrence rate of planets around low-mass stars is due to a re-distribution of the same mass into more numerous, smaller planets, we have to convert observed planet radii into planet (heavy-element) masses \citep[e.g.][]{2014MNRAS.445.3315N}. Though a one-to-one conversion is not possible due to intrinsic scatter in mass-radius based on planet composition \citep[e.g.][]{Lopez:2014er}, we will use the average mass-radius to represent our statistical ensemble. Planets smaller than $\sim1.5 ~R_\oplus$ have compositions that are consistent with rocky \citep{Rogers:2015jn,Dressing:2015je}. Larger planets contain significant gaseous atmospheres and have a mass-radius relation that is roughly linear \citep[e.g.][]{Wu:2013cp}, compared to a quadratic relationship based on a fit to solar system planets \citep{Lissauer:2011fj}. Planets up to $3 ~R_\oplus$ are significantly less dense than Neptune, with their masses set by their rocky core and their radii set by a $\sim ~1\%$ gaseous envelope \citep[e.g.][]{Wolfgang:2014uq}. For larger planets, gaseous atmospheres contain a non-negligable fraction of the planet mass. Estimates of the heavy-element mass of planets between Neptune and Jupiter sizes range from $10-100 ~M_\oplus$ \citep{Miller:2011bq}.

We use the empricial mass-radius relation from \citep[][hereafter WM14]{2014ApJ...783L...6W}, derived for planet radii smaller than $4 ~R_\oplus$, to represent the \textit{heavy-element} mass of Kepler planets. Figure \ref{f:MR} shows that this mass-radius relation happens to correctly estimate the planet \textit{heavy-element} mass for larger planets as well: for radii of $10 ~R_\oplus$, it gives a mass of $\sim23 ~M_\oplus$, surprisingly close to the median heavy-element mass of $\sim22 ~M_\oplus$ in \cite{Miller:2011bq}. We also compute planet masses using the best-fit single-power-law relation from \citep[][hereafter WRF15]{2015arXiv150407557W}. This relation predicts overall higher masses and has a steeper slope. It predicts a mass for giant planet much larger than the heavy-element mass, so we use a maximum mass of $22 ~M_\oplus$ to represent the heavy-element component, consistent wit the median heavy-element mass for giant planets in \cite{Miller:2011bq}. Because there is no seperate treatment for rocky planets, the masses of planets below $1.5 ~R_\oplus$ are overestimated with respect to WM14.

\begin{figure}
	\includegraphics[width=\linewidth]{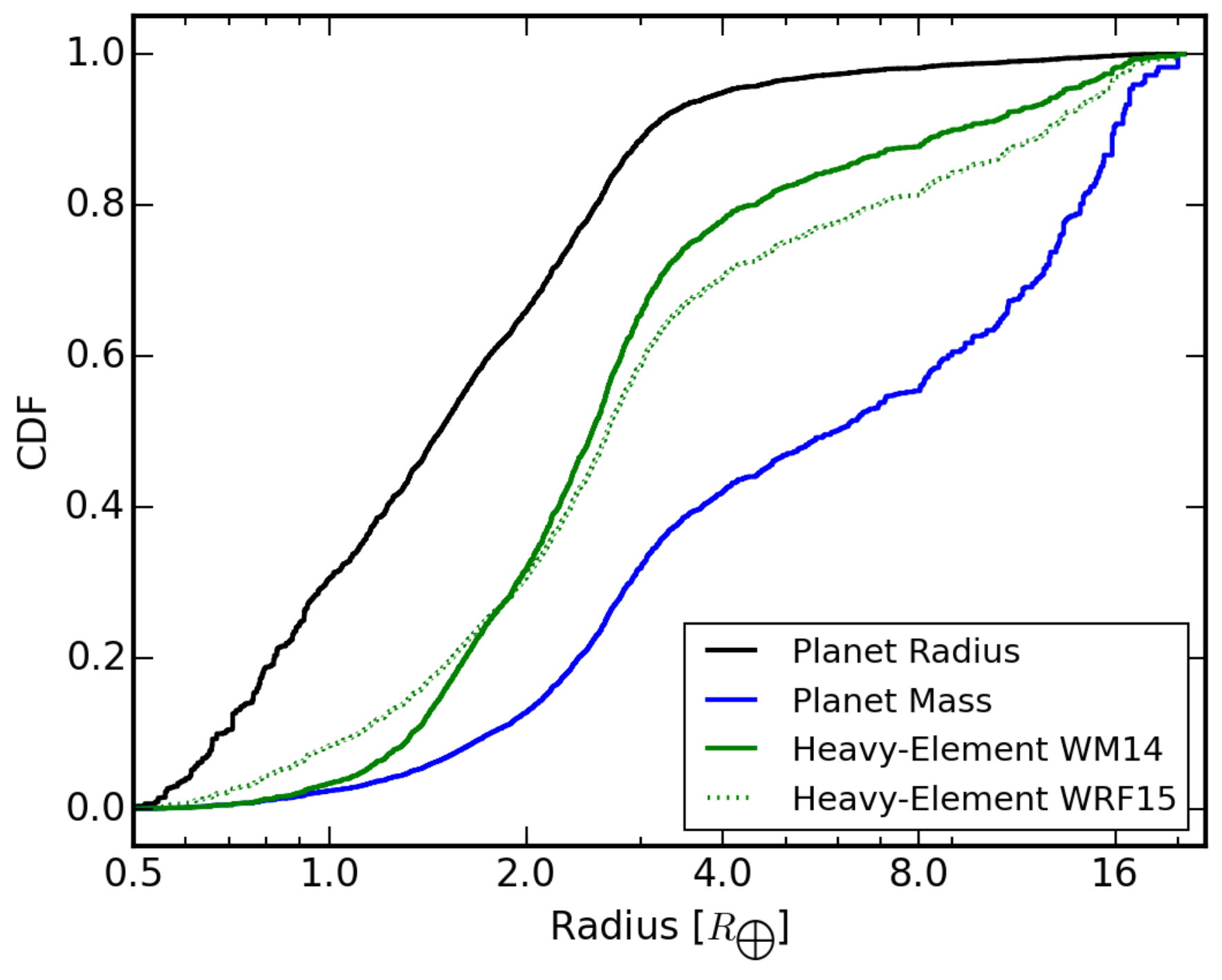}
	\caption{Cummulative planet radius distribution for orbital periods between 2 and 50 days for main-sequence stars in the Kepler sample. Despite their large numbers, rocky planets ($<1.5 R_\oplus$) do not constitute a signifcant fraction of the total heavy-element mass locked up in planets. The bulk of the heavy-element mass is locked-up in mini-Neptunes ($1.5 R_\oplus<4.0R_\oplus$).
	The green line is the cumulative heavy-element mass, based on the WR14 mass-radius relation (solid line) and on WRF15 (dotted line).
The blue line is the cumulative planet mass (gas+heavy elements), using the mass-radius relation from \cite{Lissauer:2011fj}. 
	\label{f:cdf}
	}
\end{figure}

Figure \ref{f:cdf} shows the \textit{cumulative distribution function} or CDF of planet radii, compared with an estimate of the planet \textit{heavy-element} mass as described above. The most apparent feature in the planet radius CDF is the turnover at $\sim 3 R_\oplus$, and it is clear the 90\%-95\% of the planets are smaller than $3-4 R_\oplus$. Rocky planets ($<1.5 ~R_\oplus$) constitute at least half the planet population. For comparison, the mass distribution according to quadratic mass-radius relation from \cite{Lissauer:2011fj} is shown to highlight that the total mass in planets remains dominated by the gaseous envelopes of giants. 60-70\% of the total mass is concentrated in planets larger than $3-4 R_\oplus$. 

\begin{figure}
	\includegraphics[width=\linewidth]{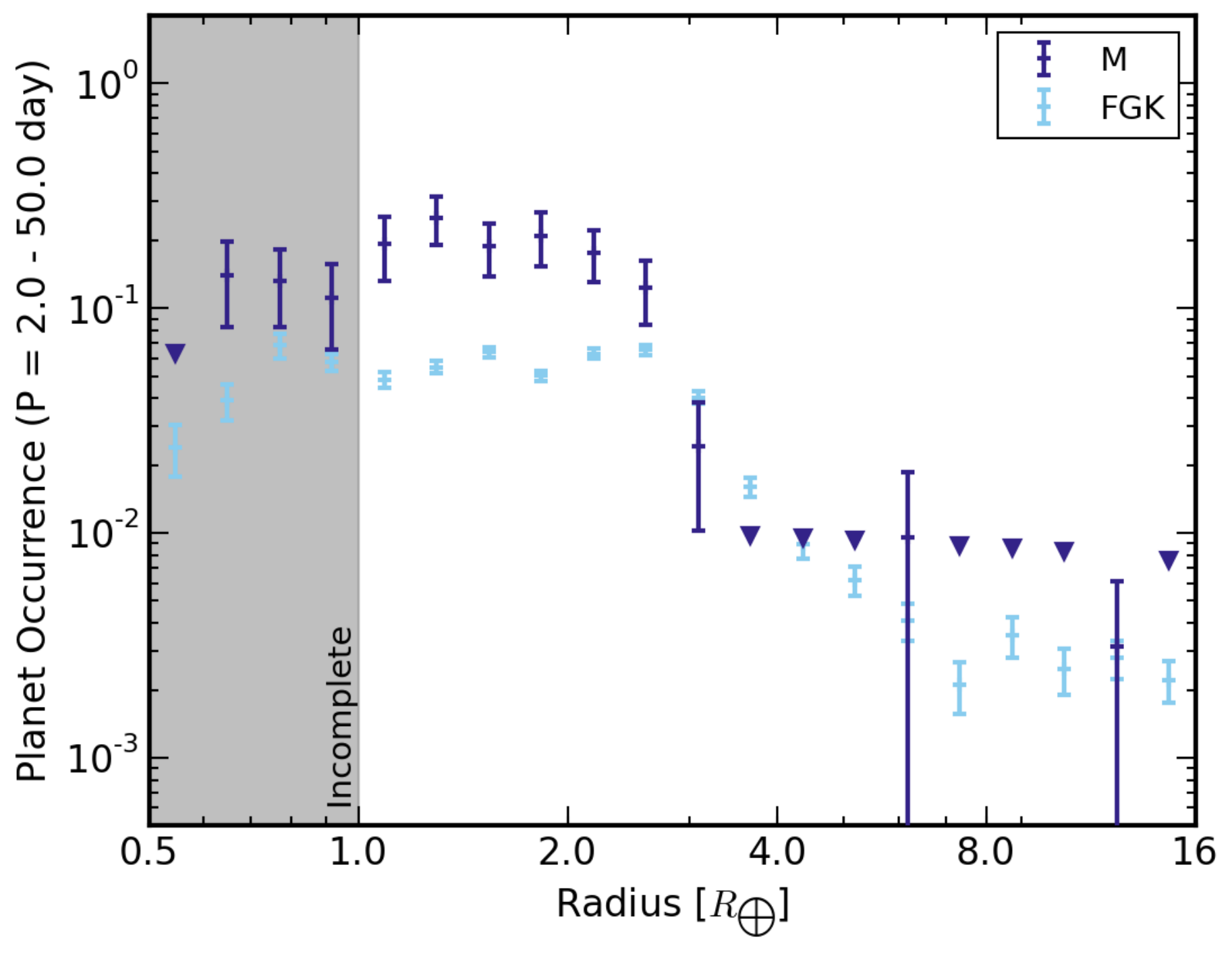}
	\includegraphics[width=\linewidth]{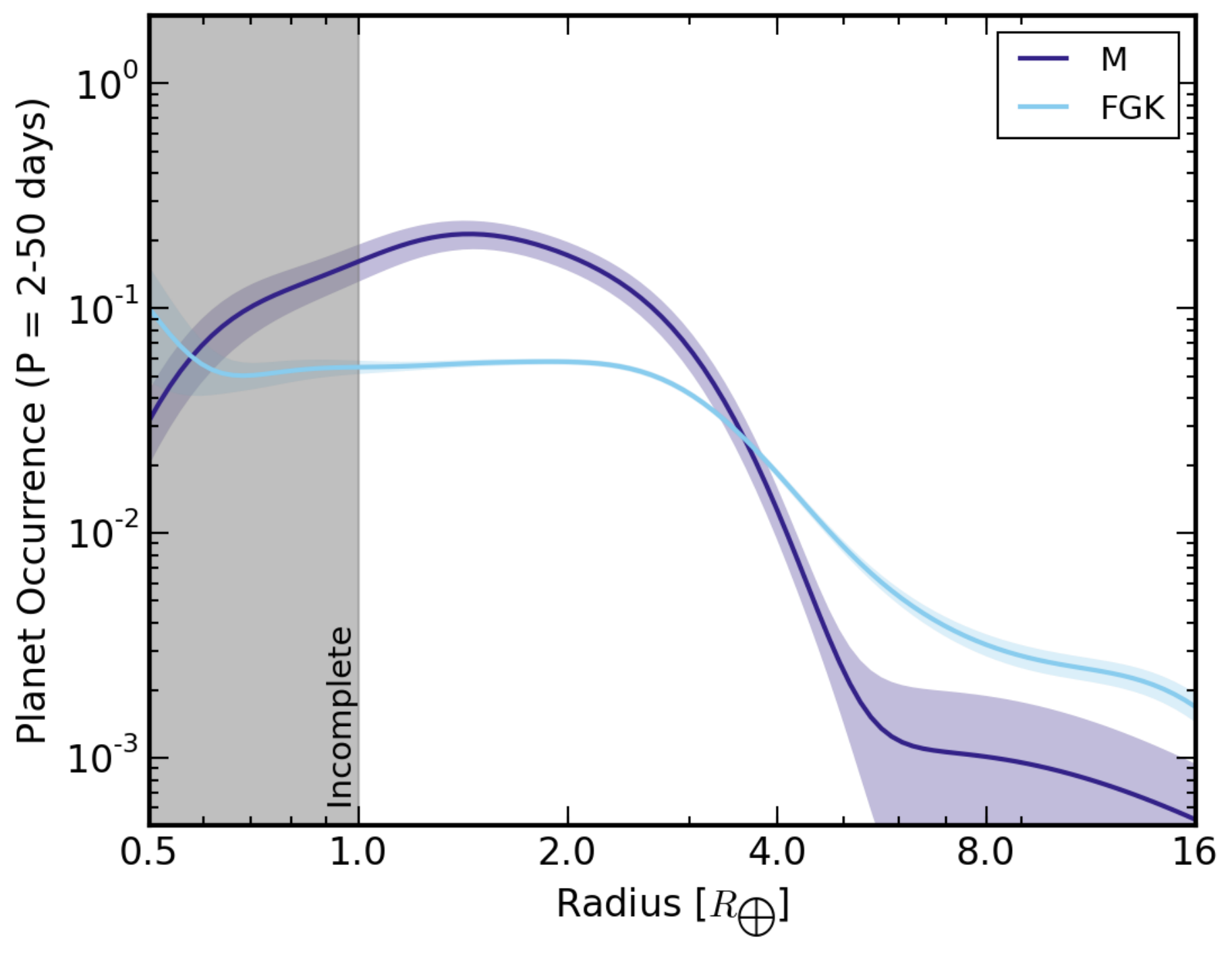}
	\caption{Planet radius distribution for orbital periods between 2 and 50 days for M dwarf stars (blue) and main-sequence FGK stars (purple). The top panel shows the binned occurrence rates {, where t}riangles denote 1-$\sigma$ upper limits. The bottom panel shows the regression curve based on gaussian kernel density estimation. The shaded region is 1-$\sigma$ confidence interval based on poisson counting statistics. Planet occurrence rates are incomplete for $<R_\oplus$. 
	\label{f:MvsFGK}
	}
\end{figure}

The distribution of heavy-elements between planets of different sizes varies. Giant planets have more heavy elements --- though as a smaller fraction of their total mass --- and the largest planets contain 20\%-30\% of the total heavy-element mass. The bulk of the heavy-element mass ($\sim$70\%) is concentrated in the cores of mini-Neptunes ($1.5-2.8 R_\oplus$), independent of the adopted mass-radius relation. About 10\% is found in super-earths ($1-1.5 R_\oplus$). The largest discrepancy between the WM14 and WRF15 relation occurs at these smallest radii, where the latter predicts double the amount of mass in planets $<1.5 R_\oplus$. This difference arises mainly because we have not included the physical mass constraint for rocky planets in the WRF15 relation. Even though the incomplete sample of sub-earth-sized planets represents 30\% of the number of planets, their masses are too small to contribute. According to the model of \cite{Zeng:2013cs}, the mass difference between a $0.7 R_\oplus$ and $1.2 R_\oplus$ planet is a factor 20. Unless the planet occurrence shows an order of magnitude increase below 1 earth radius, there is no significant contribution from small planets to the total mass. This means that Kepler has detected the bulk of the planet heavy-element mass out to a 50-day orbital period, and 80\% of this mass is located in planets with radii between $1-3 R_\oplus$.

\begin{figure}
	\includegraphics[width=\linewidth]{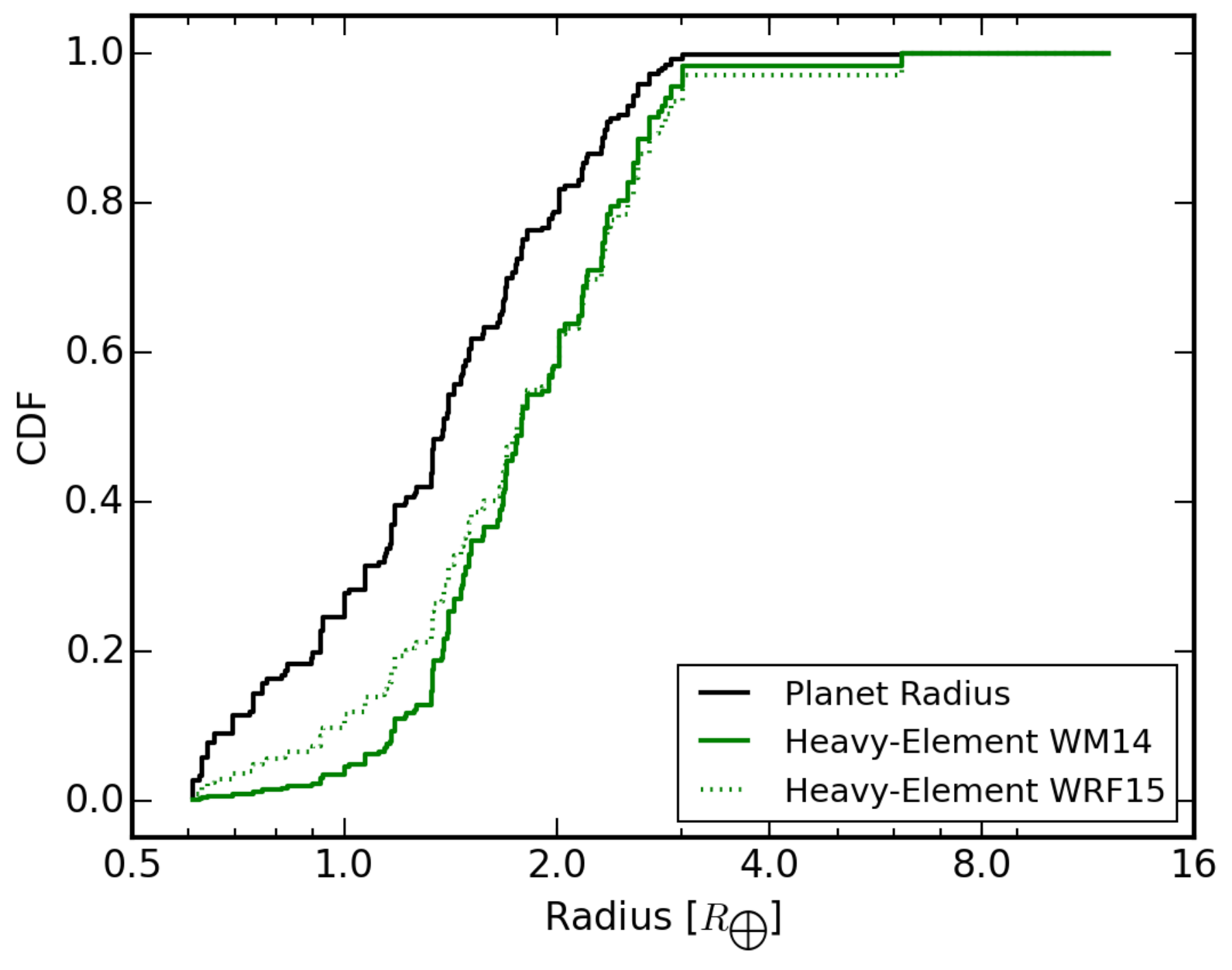}
	\caption{Same as figure \ref{f:cdf} for M dwarf stars in the Kepler sample. 95\% of the heavy-element mass is concentrated in planets smaller than $2.8-4R_\oplus$. The contribution of rocky planets ($<1.5 R_\oplus$) to the heavy-element mass is about 20\%.
	\label{f:cdfcool}
	}
\end{figure}

\section{Spectral-type dependencies}

\subsection{Higher planet occurrence rates around M stars}
The top panel of figure \ref{f:MvsFGK} shows the planet radius distribution of M dwarf stars compared to main-sequence FGK stars. Upper limits are calculated as the occurrence of a planet with a radius and orbital period corresponding to the logaritmic center of each grid cell. Because the binned version does not capture all the structure in the data, we also  show the regression curve based on variable-bandwidth kernel density estimation in the bottom panel. This approach of representing the data is similar to that of \cite{2014ApJ...791...10M}, but we use a kernel density estimator based on a maximum width of 0.1 dex and the distance to the 5th nearest neighbour. The $1-\sigma$ confidence interval is calculated by adding the contribution of each kernel in quadrature and taking the square root of the total. Both distributions show a plateau in occurrence rate between $1-2.8 R_\oplus$ and a dropoff towards larger planets. The occurrence of planets in the plateau is a factor of $\sim3.5$ higher for M stars. The occurrence rate of larger planets ($2.8-16 R_\oplus$) is a factor two lower around M stars, best visible in the regression curve, roughly consistent with the linear scaling of giant planet occurrence with stellar mass from RV surveys \citep{Johnson:2010gu}. 

We note that the low number of detected large planets around M stars is not due to a signal-to-noise bias, as Kepler is more complete for large planets than for small ones. Instead, the low number of detected KOIs is a product of the lower number of M stars surveyed combined and an intrinsic lower planet occurrence rate. In other words, if a population of planets similar to that around sunlike stars was present, it would have been detected. Instead, the ratio of small to large planets is a factor 8 higher for M stars than for FGK stars. A students t-test shows that this result is significant at the $13$ sigma level. This shows that the lack of Neptune-sized and larger planets around M stars is not due to low-number statistics. 

The CDF of the M star planets (Figure \ref{f:cdfcool}) is significantly different from that of FGK stars (Fig. \ref{f:cdf}). Planets larger than $2.8 ~R_\oplus$ do not contribute significantly to the total number or heavy-element mass of planets. 
We assume the planet mass-radius relation is the same for all spectral type sub-samples, motivated by the lack of any observed trend with stellar mass \citep{2014ApJ...783L...6W}.  95\% of the heavy-element mass around M stars is located in planets smaller than $2.8-4.0 R_\oplus$, in contrast to the 70\%-80\% in the full sample. Rocky planets amount to $\sim 10\%$ of the heavy-element mass, again implying that the contribution of planets below the detection limit to the total mass is negligible.

\begin{figure}
		\includegraphics[width=\linewidth]{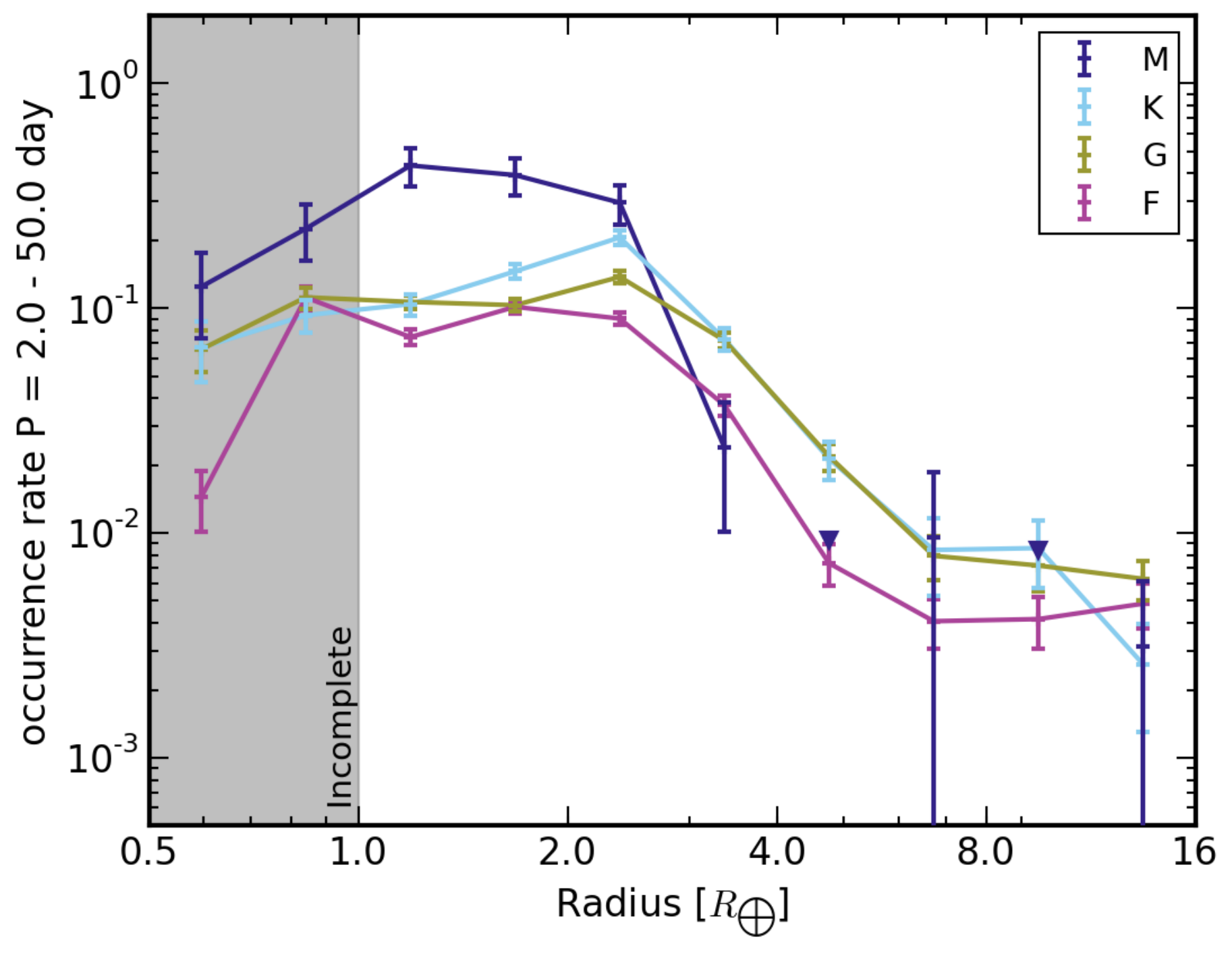}
		\includegraphics[width=\linewidth]{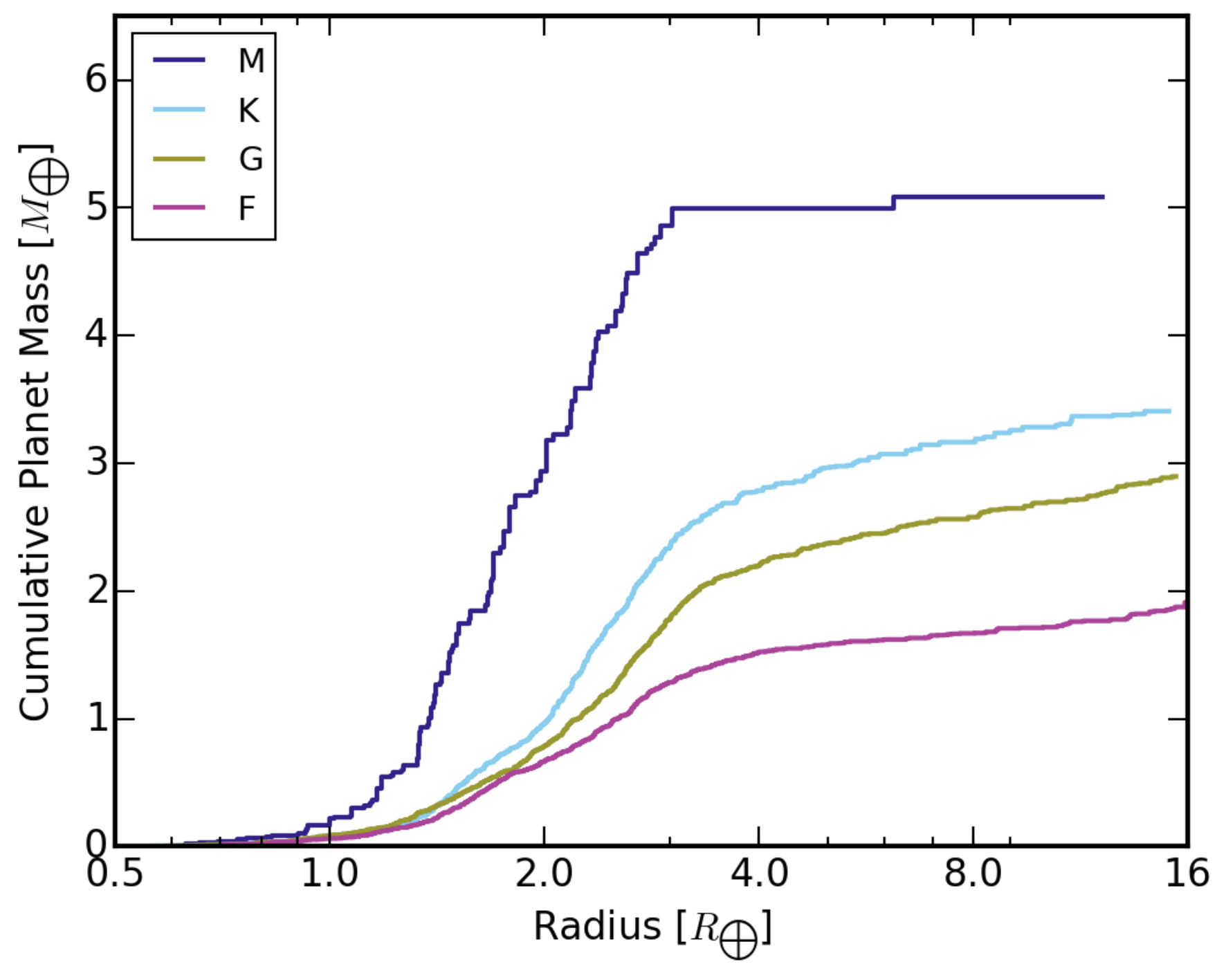}
	\caption{Planet radius distribution (left) and cumulative planet mass per star (right) for orbital periods between 2 and 50 days for M, G, K and F stars. The bins in the upper panel are twice as wide as in figure \ref{f:radius}.
	\label{f:MKGF}
	}
\end{figure}

\subsection{Differences between FGK stars}
Motivated by the finding of a higher planet occurrence rate around M stars, we investigate the spectral type dependence of the planet radius distribution between F, G, and K stars. Figure \ref{f:MKGF}, left panel shows the spectral type dependence of the binned planet radius distribution. Since the differences between FGK stars are significantly smaller than the difference with respect to M stars, we bin the planet radius to a coarser resolution. The occurrence rate in the plateau below $2.8 ~R_\oplus$ increases from $0.27\pm0.01$, to $0.35\pm0.01$, to $0.46\pm0.02$, to $1.12\pm0.12$ for F,G,K, and M stars, respectively, consistent with the increase in occurrence with spectral type of planets between $1-4 ~R_\oplus$ in \cite{Mulders:2015ja}. These difference between G stars and K and F stars are significant at the 4.5 and 5.5 sigma level, respectively. There is no such clear trend for planets larger than $2.8 ~R_\oplus$: their occurrence around K and G stars is statistically indistinguishable. Surprisingly, the occurrence of these planets around F stars is a factor of 2 lower, which is significant at the 11 sigma level.

Despite the fact that the planet radius distribution is different at each spectral type, the higher planet occurrence toward later spectral types cannot be explained by a redistribution of heavy-element mass into smaller planets. The bottom panel of Fig. \ref{f:MKGF} shows the cumulative heavy-element mass in planets for different spectral types. The planet heavy-element mass is dominated by planets smaller than ($<2.8-4 R_\oplus$) for all spectral types. The different distributions of mass among planets of different sizes are not consistent with the same amount of mass distributed among smaller planets for later spectral types. The average amount of mass locked up in M star planetary systems at $P<50$ days is a factor of 1.5, 1.7, and 2.5 higher than those in K, G and, F stars respectively.

\begin{figure}
	\includegraphics[width=\linewidth]{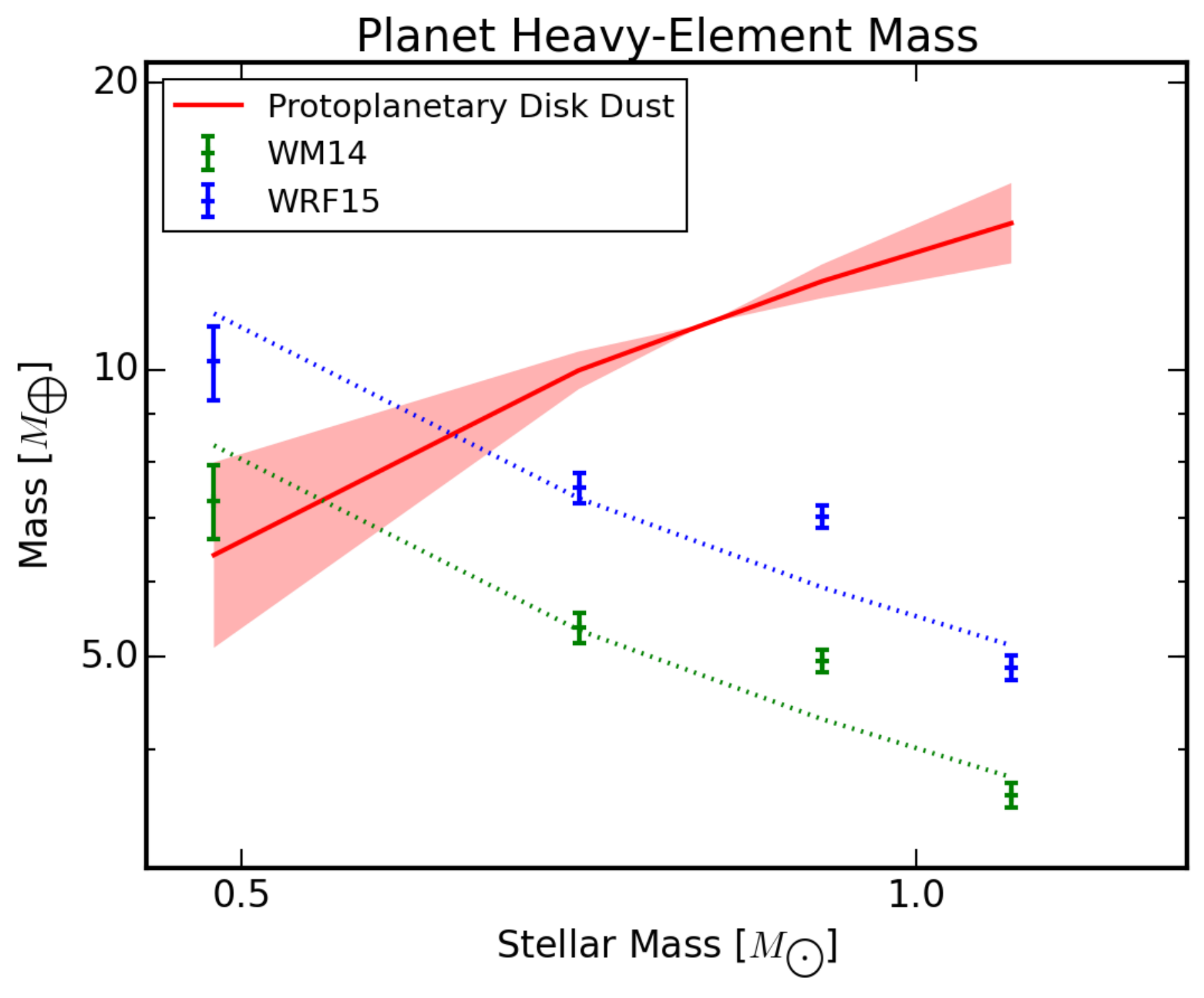}
	\caption{Average heavy-element mass locked up in planetary systems observed by Kepler versus the median stellar mass in each spectral type bin. The WRF15 mass-radius relation yields higher masses than WM14, but both relation show the same trend with stellar mass. The green and blue dotted line shows a power-law relation inversely proportional to stellar mass. The red line shows how the median dust disk mass from millimeter-wave observation scales with stellar mass, based on $M_d \approx 0.004\% ~M_\star$ by \cite{2013ApJ...771..129A}. The shaded region shows the 1$\sigma$ confidence limit on the stellar-mass-dependent scaling of disk mass with stellar mass, and does not reflect the uncertainties in the absolute values nor the dispersion at a given stellar mass.
	\label{f:disk}
	}
\end{figure}

\begin{table}
	\title{Planet Heavy-Element Mass per Spectral Type}
	
	\begin{tabular}{l l l l l}\hline\hline
	Sample & $T_{\rm eff}$ [$K$] & $M_\star$ [$M_\odot$] & $M_P$ [$M_\oplus$] & $M_d$ [$M_\oplus$]\\
	\hline
	M & 2400...3865 & 0.42 & $7.3 \pm 0.7$ & $7$ \\ 
	K & 3865...5310 & 0.73 & $5.4 \pm 0.2$ & $12$ \\ 
	G & 5310...5980 & 0.91 & $5.0 \pm 0.1$ & $15$ \\ 
	F & 5980...7320 & 1.08 & $3.6 \pm 0.1$ & $18$ \\ 
	\hline\hline\end{tabular}
	\caption{
	\ifemulate Planet Heavy-Element Mass per Spectral Type. \fi
	Average heavy-element mass of Kepler planetary systems ($M_P$) per spectral type bin, compared to median host star mass ($M_\star$) and protoplanetary disk dust masses ($M_d$) in Taurus estimated from millimeter-wave continuum observations ($M_d=4\cdot 10^{-5} M_\star$) by \cite{2013ApJ...771..129A}.}
	\label{tab:mass}
\end{table} 

\subsection{More heavy-element mass around lower-mass stars}
We calculate the average heavy-element mass in planetary systems for the largest possible region in planet radius-orbital period diagram where Kepler is complete \textit{for spectral types M to F} (See appendix \ref{a:completeness}). The heavy-element mass locked in planetary systems in this region are listed in table \ref{tab:mass} and shown in Figure \ref{f:disk} for each spectral type sub-sample. This mass increases from $3.6 M_\oplus$ in F stars to  $7.3 ~M_\oplus$ in M stars and scales roughly proportional to the inverse of the median stellar mass in each spectral type bin which is listed in the same table. 
The inclusion of the 50-150 day region increases the total planet mass with respect to that of Figure \ref{f:MKGF}, but we have tested that it does not change the trend with spectral type.
For comparison, the estimated dust mass in protoplanetary disks shows the opposite trend with stellar mass \citep{2013ApJ...773..168M,2013ApJ...771..129A}. Although dust disk masses are measured from and dominated by a region further away from the star, the scaling law with stellar mass is often assumed to be relevant for formation of close-in planets \citep[e.g.][]{2007ApJ...669..606R}.  The agreement between dust mass and planet heavy-element mass for M stars is likely coincidental: surveys are insensitive to objects larger than a few centimeters, and it is likely that a larger reservoir of planetary building blocks is present in these disks \citep[e.g.][]{2014MNRAS.445.3315N}. 

This trend is robust against the various assumptions that go into deriving it. Our estimates for the M star planet occurrence rates are lower than those in \cite{2015ApJ...807...45D}, most likely due to our overestimation of the pipeline completeness for M stars versus FGK stars. However, a higher planet occurrence rate for M dwarfs would only strengthen the observed trend. Adopting a different mass-radius relation increases the total estimated heavy-element mass, but does not affect the observed trend with spectral type (See Fig.  \ref{f:disk}). The only factor that could significantly change this result is if the mass-radius relation is dependent on spectral type. There are currently no indications that this is the case \citep{2014ApJ...783L...6W}, though only few measurements of M star planet masses and radii are avaiable. To be consistent with a linear disk-stellar mass scaling, planets around M stars of sizes $\sim 1-3 ~R_\oplus$ need to be 5-10 times less dense than around sun-like stars. Though we do no deem this scenario to be very likely, it would have interesting implications for planet formation theories.

\section{Discussion}
The increase in the average mass of planetary systems toward low mass stars found in the \textit{Kepler} data goes against the widely adopted assumption that planet masses scale with total disk mass and hence should \textit{decrease} with stellar mass. In this section we will compare these results to the predictions of planet formation in-situ (\S \ref{s:insitu}) and through migration of fully formed planets (\S \ref{s:migrate}). We also show that we can exclude binarity (\S \ref{s:binary}) and long-period giant planets (\S \ref{s:giant}) as alternate explanations for the increased occurrence rates, and propose a scenario to explain these results (\S \ref{s:bb}, \S \ref{s:eff}).

\subsection{Comparison with in-situ formation}\label{s:insitu}
\cite{2006ApJ...642.1131K} explored how changes in the disk surface density affect terrestrial planet formation between 0.5 and 1.5AU around a solar-mass star using $N$-body simulations. They fit a relation between the disk surface density $\Sigma$ and the number of planets, $n$, and their masses, $M_P$, that can form in situ in such a disk. Assuming the surface density is proportional to disk mass, the number of planets scales as $n \propto \Sigma^{-0.15}$. If we assume surface density is the only factor determining the number and mass of planets, the stellar mass range of a factor 2 probed in the Kepler data predicts an increase in the occurrence of planets of 10\%, significantly smaller than the factor 3.5 derived in this work for $1-2.8 ~R_\oplus$ planets. The predicted planet mass scale slightly steeper than linear, $M_P \propto \Sigma^{1.1}$. Factoring in the planets mass-radius distribution, $R_P \propto M_P^{0.93}$, a systematic shift in the planet radius distribution of a factor of $\sim 2$ between M stars and FGK stars is predicted, yet such a trend is not present in the Kepler data (see Figure \ref{f:MvsFGK}).

The scaling laws from \cite{2006ApJ...642.1131K} do not take into account differences in stellar mass, which may lead to a different scaling relation between surface density and number of planets. We also compare our result with the N-body simulations of \cite{Ciesla:2015ha}, who simulate the formation of terrestrial planets around stars of various masses. They assume the total surface surface density of planetary building blocks scales with stellar mass. Although these simulations cover larger orbital periods and smaller planet radii than accessible to Kepler, they do provide insight in how the number of planets scales with stellar mass and surface density in the context of in-situ formation. Between 0.5 and 1.5 AU, the simulations for a sun-like star contain 14 planets. The corresponding \textit{orbital period} range for the $0.4 ~M_\odot$ star, representative of the M star sample, contains 18 planets. Again, the decrease in surface density does not increase the number of small planets formed in situ by a large enough number ($1.3$ versus $3.5$) to explain the increased occurrence of planets around low-mass stars. In addition, these planets are 3-5 times less massive, a feature not seen in the planet radius distribution. Therefore, existing in-site models do not appear to predict the general trend in planet occurrence rates and stellar mass found in our study.

\subsection{Comparison with planet migration}\label{s:migrate}
\cite{2013ApJ...764..105S} proposed the convergent migration of fully-formed planets from beyond the ice line to explain the presence of close-in super-earths around M stars. In this scenario, the planet's mass and occurrence rate at short orbits are determined by the disk properties at larger distances from the star. Owing to their figure 6, isolation masses similar to Kepler planets are reached outside of $\sim 10$ au. The isolation mass, $M_{\rm ISO}$, scales with stellar mass, surface density, and distance from the star, $a$, as $M_{\rm ISO} \propto M_\star^{-1/2} ~\Sigma^{3/2} ~a^3$. The planet's Hill sphere, a measure of the separation between planets, scales as $R_H \propto a \sqrt[3]{M_{\rm ISO}/M_\star}$. Factoring in a linear scaling between surface density and stellar mass based on millimeter-wave observations, the planet mass at fixed distance from the star scales with stellar mass as $M_{\rm ISO} \propto M_\star$. The Hill sphere is independent of stellar mass, and the same number of isolation masses are formed. This model therefore predicts the same number of planets for low-mass stars, but smaller in size.
This prediction is similar to that of the in-situ formation models, and is inconsistent with the increased planet occurrence towards lower mass stars and the lack of a strong trend in the planet radius distribution function with stellar mass.

\subsection{Fewer binary companions around later spectral types}\label{s:binary}
If binary companions inhibit planet formation, a lower binarity fraction towards later spectral type could explain the rising trend in planet occurrence. \cite{Mulders:2015ja} discussed the impact of binaries on planet formation in the context of in situ formation, but found the effects to be insufficient to explain the trend in occurrence rate. If close-in planets form from material farther out in the disk and migrated to their current locations, the possible disruptive impact of a binary companion is larger. Non-redundant aperture mask imaging of Kepler planet hosts show that they indeed lack short period ($<50 au$) binary companions, suggesting that planet formation is indeed inhibited by the presence of these companions (Kraus et al., ApJ Submitted). 

Binaries with separations less than $50$ au represent a significant fraction of the field star population. According to \cite{2010ApJS..190....1R}, $50\%$ of G stars and $35\%$ of M stars have binary companions, of which approximately half interior to $50$ au. Assuming a $100\%$ planet formation efficiency for wider binaries, planet occurrence rate could increase by a factor $(1-0.175)/(1-0.25)= 1.1$ around M dwarfs, i.e. nowhere close to the factor 3.5 difference in planet occurrence rate. Even in the extreme case where binary companions would inhibit planet formation at all separations, the planet occurrence rate around M dwarfs would only increase by a factor of 1.3 with respect to that around G stars and could only partly explain the factor of 3.5 that is observed.

\subsection{Fewer long-period giant planets around M dwarfs}\label{s:giant}
\cite{Izidoro:2015co} proposed that giant planets formed at the snow line act as dynamical barriers to inward migrating super-earths and mini-Neptunes. Hence, the occurrence rates derived in this paper would be anti-correlated with those of long period giants. A similar correlation would be expected if subsequent inward and outward migration of giant planets destroys close-in super-Earths \citep{Batygin:2015dh}. In the core accretion model, low-mass disks around low-mass stars would be less likely to form giant planets \citep[e.g.][]{Laughlin:2004hf}, which might explain the overabundance of super-earths around low-mass stars. \cite{Johnson:2010gu} find that the occurrence rate of giant planets increases roughly linear with stellar mass, from 3\% in M stars, $\sim 8.5\%$ in FGK stars, to 14\% around A stars. These percentages are not high enough to explain the observed trend in Kepler planet occurrence, but it should be noted that current radial velocity surveys probe a limited range in baselines (3-10 years) and are likely incomplete in planet mass at longer periods. More complete estimates of the giant planet fraction find $f_M \sim 15\%$ for M stars including RV trends, \citep{Montet:2014fa} and $f_G \sim 20 \%$ for the population down to Saturn sizes and out to 20 au \citep{Cumming:2008hg}. 

The required giant planet fractions to explain the trend in Kepler occurrence rates are much higher. A giant planet occurrence of $f_G \sim 70\%$ around G stars (and corresponding $f_M \sim 30\%$ in Mstars) is neccesary to reach $(1-f_M)/(1-f_G) \approx 2$. These numbers should be seen as lower limits, as giant planets may not stop all inward-migrating super-earths \citep{Izidoro:2015co}. Hence, the presence of long-period giants may only explain part of the increase in planet occurrence rates towards lower stellar masses.

\subsection{Efficient redistribution of heavy-elements}\label{s:bb}
If large-scale inward migration of solid material is more effective towards low-mass stars, an even larger population of planets must be present around higher mass stars at larger distances. Kepler only probes the planet population inside of roughly an au, while radial velocity surveys that probe larger distances from the star are mostly sensitive to giant planets. If smaller planets are sequestered at radii larger than $\sim 1$ au, they would escape detection in both samples. Micro-lensing surveys indicate that a population of super-earths ($5-10 ~M_\oplus$) and Neptunes ($10-30 ~M_\oplus$) are present around approximately half the (MGK) stars in the galaxy between $0.5-10$ au \citep{Cassan:2012in}. This population presents a comparable amount of (heavy-element) mass in planetary systems as the Kepler planet population. There is no information on the stellar-mass dependence of this population, as most host stars remain unidentified. If the average planet mass and occurrence of this population correlates positively with stellar mass, it could represent the missing planet mass not seen at short orbits by Kepler. However, the planet mass and occurrence from micro-lensing surveys are estimated assuming no such scaling with stellar mass, and a stellar-mass dependency will alter the estimated planet mass and occurrence rates of this population.

There are two very efficient mechanisms for inward migration: radial drift of dust grains \citep{1977MNRAS.180...57W}, and type-I migration of (proto)planets \citep[e.g.][]{Goldreich:1979iq,1979MNRAS.186..799L}. If either of these mechanisms is responsible for increasing the heavy-element mass in the inner regions to create the population of super-earths and mini-Neptunes at short orbital periods, a stellar-mass dependency in the efficiency of this mechanism could explain the trends identified in this work. The efficiency of the first mechanism, radial drift of dust grains, was explored by \cite{2013A&A...554A..95P}. The authors simulated the growth and drift of dust grains in brown dwarf disks, and find that radial drift is more efficient towards low-mass stars. If these inward drifting solids can be maintained in the inner disk by growing them into larger objects that stop drifting, for example by enhanced sticking \citep{Boley:2014ck}, gravitational collapse \citep{Chatterjee:2014hw, Chatterjee:2015eo}, or pebble accretion onto pre-existing cores \citep{2012A&A...544A..32L,2015arXiv150601666M}, such a scenario would lead to an increased amount of heavy elements in the inner disk available for planet formation in low-mass stars. This mechanism would also deliver large amounts of volatiles to the inner disk when inward migrating solids cross the snow line and sublimate. It is not clear if this is consistent with the lack of observed water vapor in low-mass stars \citep{Pascucci:2013ei}.

The efficiency of the other mechanism, type-I migration, can be estimated from the planet migration time-scale and the extent to which planets are trapped at short orbital periods. According to Eq. 70 from \cite{Tanaka:2002ev} the planet migration time-scale, $\tau$ scales with stellar mass, disk temperature, $T$ and surface density as: $\tau \propto M_\star^{1/2} ~\Sigma ~T$. Assuming a disk temperature of $T \propto L_{\rm PMS}^{1/4} \propto M_\star^{1/2}$ during the pre-main sequence and a surface density proportional to stellar mass yields a migration time-scale, $\tau \propto M_\star^0$, i.e. independent of stellar mass. Hence it is not clear why migration of planetary building blocks would be more efficient around low-mass stars. 
Alternatively, lower-mass stars may be more efficient in halting migrating proto-planets at short orbital periods. Planet migration traps are a crucial ingredient in planet migration theories \citep[e.g.][]{Ida:2010ki}. The inner edge of the gas disk presents a powerful planet trap \citep[e.g.][]{Masset:2006dv}, and our previous results indicate that the location of this trap is stellar-mass dependent \citep{Mulders:2015ja}. The 
results of the current paper indicate that also the \textit{efficiency} of this trap may be stellar-mass dependent. Planet traps also exists further out in the disk, associated with heat and surface density transitions in the disk \citep[e.g.][]{2012ApJ...760..117H}. The relative importance of these traps changes with stellar mass, and may result in more close-in planets around low-mass stars \cite{2013ApJ...778...78H}. 

\subsection{A higher planet formation efficiency}\label{s:eff}
The higher occurrence rate for close-in planets may simply point to a higher planet formation efficiency around lower mass stars, as also suggested by \cite{Muirhead:2015gw} based on the high occurrence of compact multiples around mid-M dwarfs. Why lower-mass stars would be more efficient in converting their solid mass into planetary systems remains an open question. Perhaps the higher planet occurrence may be linked to longer disk lifetimes around low-mass stars \citep[e.g.][]{2005AJ....129.1049C,2012ApJ...758...31L}, allowing for more time for planets to form or migrate.

\section{Conclusion}
We derive the planet radius distribution for main-sequence stars of spectral types M to F probed by the \textit{Kepler} spacecraft, and make an inventory of the average \textit{heavy-element} mass locked up in planetary systems at short orbital periods. We find that:

\begin{itemize}
\item The average number of small planets ($1.0-2.8 ~R_\oplus$) per M star is a factor 3.5 higher than that per FGK star at orbital periods $P<50$ days. The number of larger planets (``Neptunes'') is factor of 2 lower. Aside from the lack of Neptunes around M stars, there is no systematic trend in the planet radius distribution between F, G, and K stars that points to smaller planets forming around lower-mass stars. 

\item Taking into account the observed mass-radius relation for exoplanets, we show that higher occurrence rates around M stars are not the result of a redistribution of the same heavy-element mass into more numerous, smaller planets. M stars have, on average, a larger amount of heavy-element mass locked up in planetary systems at short orbital periods than FGK stars.

\item The majority of the \textit{heavy-element} mass in short-period planets around FGK stars is locked up into mini-Neptunes ($1.5-4.0 ~R_\oplus$, $\sim 70\%$), with smaller fractions in rocky planets ($<1.5 ~R_\oplus$, $\sim 10\%$) and Neptunes ($>4 ~R_\oplus$, $\sim 10\%$). In M stars planetary systems, close to $100\%$ of the (heavy-element) mass is located in planets smaller than $4.0 ~R_\oplus$, with $\sim 20\%$ in rocky planets smaller than $1.5 ~R_\oplus$.

\item Planets below the detection limit do not constitute a significant amount of (heavy-element) mass for $P < 50$ days, and Kepler has detected the bulk of the planet (heavy-element) mass out to 150 day orbital periods. In the region where Kepler is complete for all spectral types, the average heavy-element mass in planetary systems increase from $3.6 \pm 0.1 ~M_\oplus$ in F stars to $5.0 \pm 0.1 ~M_\oplus$ in G stars, $5.4 \pm 0.2 ~M_\oplus$ in K stars, and $7.3 ~M_\oplus \pm 0.7$ in M stars. 

\end{itemize}

The roughly linear anticorrelation between heavy-element mass in planetary systems and stellar mass contrasts with the observed linear correlation between disk mass and stellar mass. This feature is currently not predicted by planet formation theories, either in situ or by migration. Such a feature may be explained by enhanced radial drift of solids of more efficient type-I migration of planetary building blocks towards lower-mass stars.

\ifemulate
	\vspace{1em}{\it Acknowledgments:}\\
\else
	\acknowledgments
\fi
This paper includes data collected by the Kepler mission. Funding for the Kepler mission is provided by the NASA Science Mission directorate.
We thank the referee for the careful and constructive review.
We would like to thank Fred Ciesla for providing comments on the manuscript, and Jessie Christiansen and Adam Kraus for sharing preprints of their papers.
We would also like to thank Dean Billheimer and his students for statistical advice through the Statistical Consulting course.
This material is based upon work supported by the National Aeronautics and Space Administration under Agreement No. NNX15AD94G for the program “Earths in Other Solar Systems”. 
The results reported herein benefitted from collaborations and/or information exchange within NASA’s Nexus for Exoplanet System Science (NExSS) research coordination network sponsored by NASA’s Science Mission Directorate.

\appendix
\begin{figure*}
	\includegraphics[width=\fighalfwidth\linewidth]{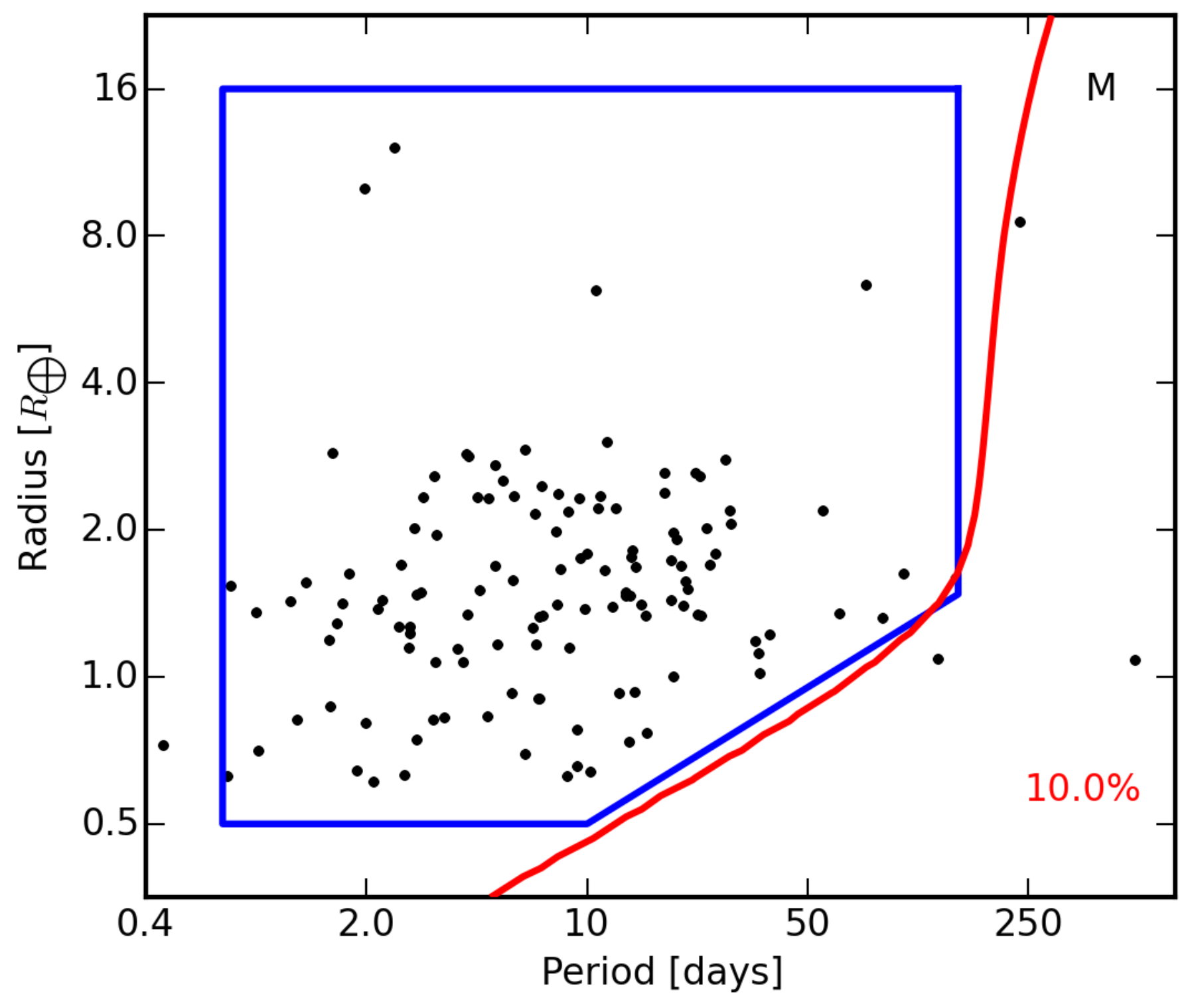}
	\includegraphics[width=\fighalfwidth\linewidth]{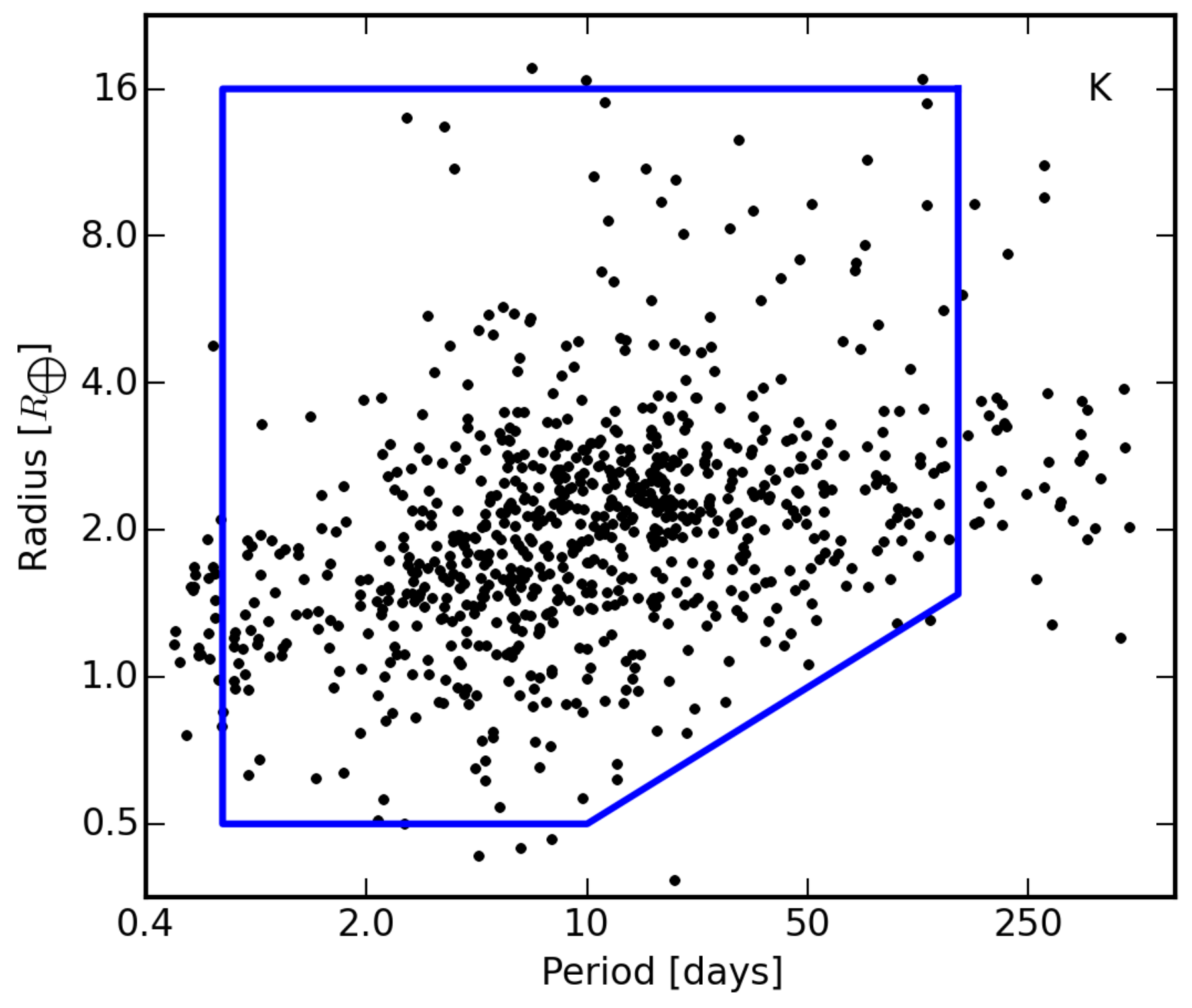}\\
	\includegraphics[width=\fighalfwidth\linewidth]{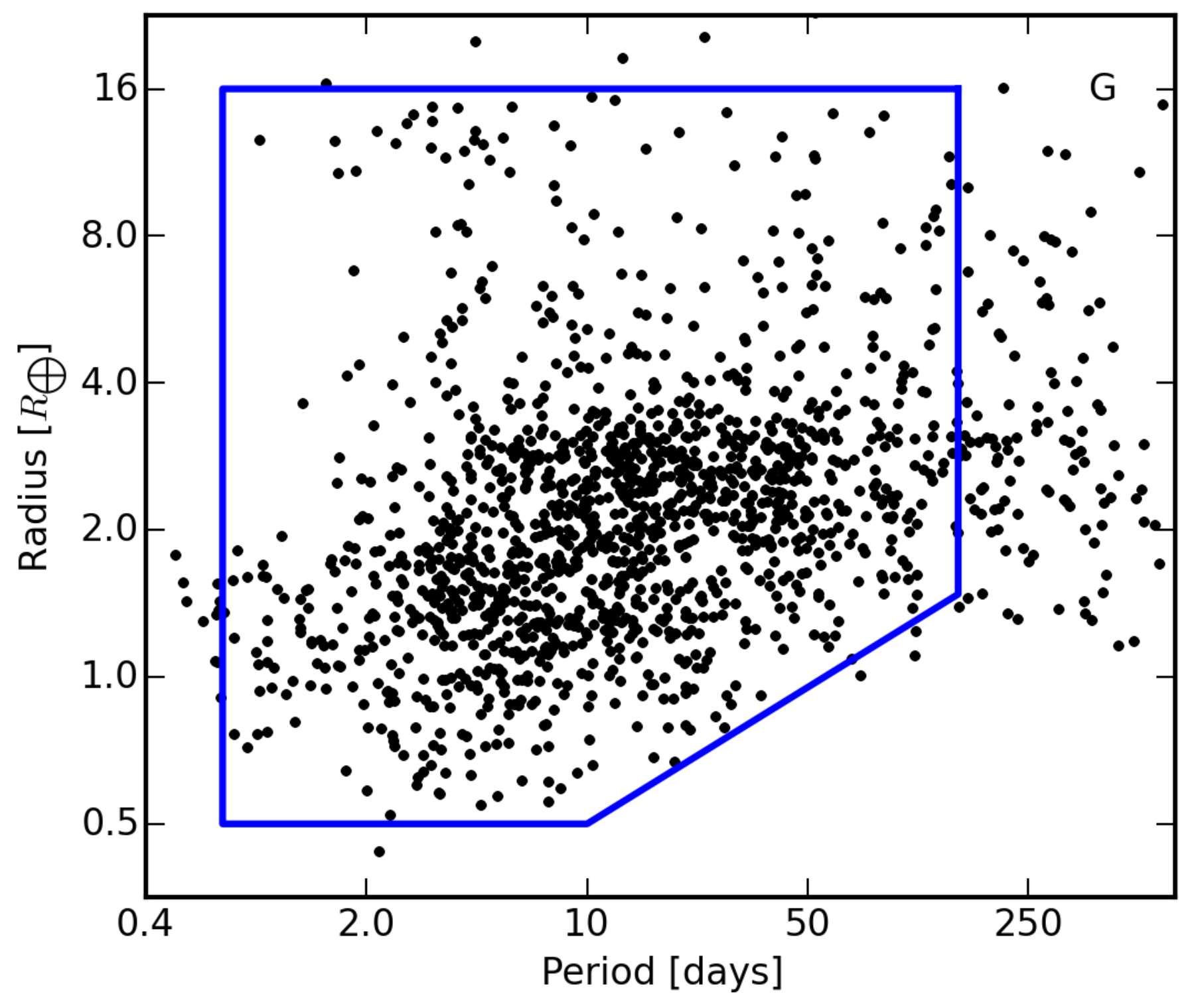}
	\includegraphics[width=\fighalfwidth\linewidth]{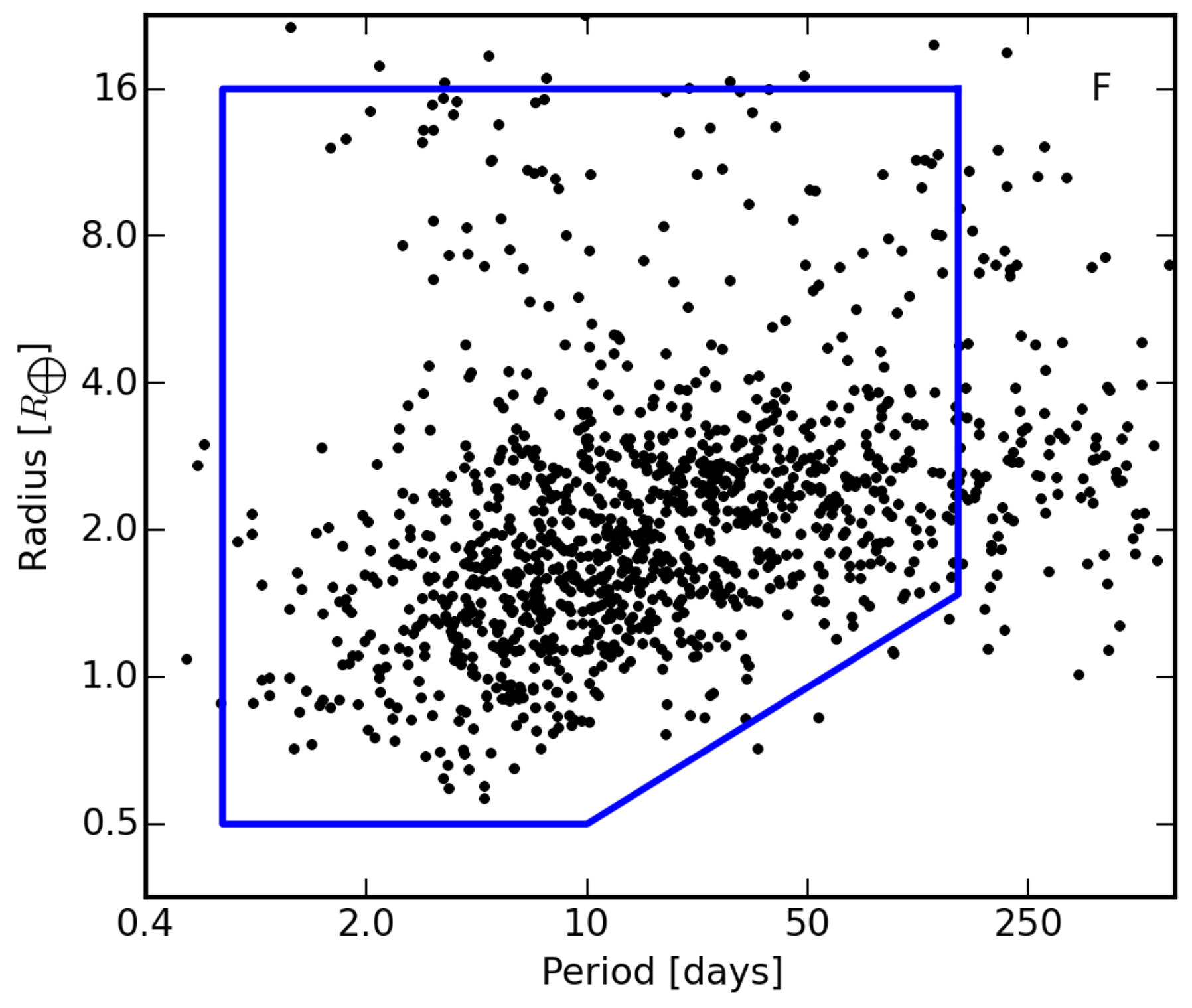}
	\caption{
	Completeness limit overlaid on the KOI distribution for spectral type sub-samples M (top left), K (top right), G (bottom left, and F (bottom right). The red contour in the M star panel indicates the location out to which an occurrence rate of 10\% can be detected.
	\label{f:complete}
	}
\end{figure*}

\section{Completeness per spectral type}\label{a:completeness}
Figure \ref{f:complete} shows the distribution of KOIs for spectral type sub-sample M to F. The region where Kepler is complete, indicated in blue, is limited by the completeness of the M star sample, which has the lowest number of stars. The red contour indicates the location where the expected detection frequency is one per bin for an occurrence rate of 10\%, which is equivalent to an occurrence rate of 3 per logaritmic area unit. The lower limit of $0.5 ~R_\oplus$ is chosen because there are very few detection below. The M star sample limits the completeness to $P<150$ days, $R_P<16 ~R_\oplus$, and excludes a wedge from $(P,R_P)=(15,0.7)$ to $(150,1.4)$. We place the inner edge at an orbital period of $0.7$ days. As shown in Figures \ref{f:cdf} and \ref{f:cdfcool}, the excluded planets in the wedge contains at most $10\%$ of the heavy-element mass at those orbital periods because they are smaller than $<1.5 ~R_\oplus$. Varying the inner, lower, and upper boundary changes the total mass by only a few percent, and we have verified that this does not significantly change the trends with spectral type.

\ifemulate
	\ifastroph
		\bibliography{paper.bbl}
	\else	
		\bibliography{papers3}
	\fi
\else	
	\bibliography{paper.bbl}

\begin{thebibliography}{}
\expandafter\ifx\csname natexlab\endcsname\relax\def\natexlab#1{#1}\fi

\bibitem[{Adibekyan {et~al.}(2015)Adibekyan, Figueira, \&
  Santos}]{Adibekyan:2015vb}
Adibekyan, V., Figueira, P., \& Santos, N.~C. 2015, eprint arXiv:1509.02429,
  1509.02429

\bibitem[{Alibert {et~al.}(2013)Alibert, Carron, Fortier, Pfyffer, Benz,
  Mordasini, \& Swoboda}]{2013A&A...558A.109A}
Alibert, Y., Carron, F., Fortier, A., {et~al.} 2013, Astronomy {\&}
  Astrophysics, 558, A109

\bibitem[{Andrews {et~al.}(2013)Andrews, Rosenfeld, Kraus, \&
  Wilner}]{2013ApJ...771..129A}
Andrews, S., Rosenfeld, K.~A., Kraus, A.~L., \& Wilner, D.~J. 2013, The
  Astrophysical Journal, 771, 129

\bibitem[{Batygin \& Laughlin(2015)}]{Batygin:2015dh}
Batygin, K., \& Laughlin, G. 2015, Proceedings of the National Academy of
  Science, 112, 4214

\bibitem[{Boley {et~al.}(2014)Boley, Morris, \& Ford}]{Boley:2014ck}
Boley, A.~C., Morris, M.~A., \& Ford, E.~B. 2014, Astrophysical Journal, 792,
  L27

\bibitem[{Borucki {et~al.}(2011)Borucki, Koch, Basri, Batalha, Boss, Brown,
  Caldwell, Christensen-Dalsgaard, Cochran, DeVore, Dunham, Dupree, Gautier,
  Geary, Gilliland, Gould, Howell, Jenkins, Kjeldsen, Latham, Lissauer, Marcy,
  Monet, Sasselov, Tarter, Charbonneau, Doyle, Ford, Fortney, Holman, Seager,
  Steffen, Welsh, Allen, Bryson, Buchhave, Chandrasekaran, Christiansen,
  Ciardi, Clarke, Dotson, Endl, Fischer, Fressin, Haas, Horch, Howard,
  Isaacson, Kolodziejczak, Li, MacQueen, Meibom, Prsa, Quintana, Rowe, Sherry,
  Tenenbaum, Torres, Twicken, Van~Cleve, Walkowicz, \& Wu}]{Borucki:2011cp}
Borucki, W.~J., Koch, D.~G., Basri, G., {et~al.} 2011, Astrophysical Journal,
  728, 117

\bibitem[{Buchhave \& Latham(2015)}]{Buchhave:2015cg}
Buchhave, L.~A., \& Latham, D.~W. 2015, Astrophysical Journal, 808, 187

\bibitem[{Buchhave {et~al.}(2012)Buchhave, Batalha, Latham, Johansen, Bizzarro,
  Torres, Rowe, Borucki, Brugamyer, Caldwell, Bryson, Ciardi, Cochran, Endl,
  Esquerdo, Ford, Geary, Gilliland, Hansen, Isaacson, Laird, Lucas, Marcy,
  Morse, Robertson, Shporer, Stefanik, Still, \& Quinn}]{2012Natur.486..375B}
Buchhave, L.~A., Batalha, N., Latham, D.~W., {et~al.} 2012, Nature, 486, 375

\bibitem[{Burke {et~al.}(2014)Burke, Bryson, Mullally, Rowe, Christiansen,
  Thompson, Coughlin, Haas, Batalha, Caldwell, Jenkins, Still, Barclay,
  Borucki, Chaplin, Ciardi, Clarke, Cochran, Demory, Esquerdo, Gautier,
  Gilliland, Girouard, Havel, Henze, Howell, Huber, Latham, Li, Morehead,
  Morton, Pepper, Quintana, Ragozzine, Seader, Shah, Shporer, Tenenbaum,
  Twicken, \& Wolfgang}]{2014ApJS..210...19B}
Burke, C.~J., Bryson, S.~T., Mullally, F., {et~al.} 2014, The Astrophysical
  Journal Supplement, 210, 19

\bibitem[{Burke {et~al.}(2015)Burke, Christiansen, Mullally, Seader, Huber,
  Rowe, Coughlin, Thompson, Catanzarite, Clarke, Morton, Caldwell, Bryson,
  Haas, Batalha, Jenkins, Tenenbaum, Twicken, Li, Quintana, Barclay, Henze,
  Borucki, Howell, \& Still}]{2015ApJ...809....8B}
Burke, C.~J., Christiansen, J.~L., Mullally, F., {et~al.} 2015, The
  Astrophysical Journal, 809, 8

\bibitem[{Carpenter {et~al.}(2005)Carpenter, Wolf, Schreyer, Launhardt, \&
  Henning}]{2005AJ....129.1049C}
Carpenter, J.~M., Wolf, S., Schreyer, K., Launhardt, R., \& Henning, T. 2005,
  The Astronomical Journal, 129, 1049

\bibitem[{Cassan {et~al.}(2012)Cassan, Kubas, Beaulieu, Dominik, Horne,
  Greenhill, Wambsganss, Menzies, Williams, J{\o}rgensen, Udalski, Bennett,
  Albrow, Batista, Brillant, Caldwell, Cole, Coutures, Cook, Dieters, Prester,
  Donatowicz, Fouqu{\'e}, Hill, Kains, Kane, Marquette, Martin, Pollard, Sahu,
  Vinter, Warren, Watson, Zub, Sumi, Szyma{\'{n}}ski, Kubiak, Poleski,
  Soszynski, Ulaczyk, Pietrzy{\'{n}}ski, \& Wyrzykowski}]{Cassan:2012in}
Cassan, A., Kubas, D., Beaulieu, J.~P., {et~al.} 2012, Nature, 481, 167

\bibitem[{Chatterjee \& Tan(2014)}]{Chatterjee:2014hw}
Chatterjee, S., \& Tan, J.~C. 2014, Astrophysical Journal, 780, 53

\bibitem[{Chatterjee \& Tan(2015)}]{Chatterjee:2015eo}
---. 2015, Astrophysical Journal, 798, L32

\bibitem[{Chiang \& Laughlin(2013)}]{2013MNRAS.431.3444C}
Chiang, E., \& Laughlin, G. 2013, Monthly Notices of the Royal Astronomical
  Society, 431, 3444

\bibitem[{Christiansen {et~al.}(2012)Christiansen, Jenkins, Caldwell, Burke,
  Tenenbaum, Seader, Thompson, Barclay, Clarke, Li, Smith, Stumpe, Twicken, \&
  Van~Cleve}]{Christiansen:2012bz}
Christiansen, J.~L., Jenkins, J.~M., Caldwell, D.~A., {et~al.} 2012,
  Publications of the Astronomical Society of the Pacific, 124, 1279

\bibitem[{Christiansen {et~al.}(2015)Christiansen, Clarke, Burke, Seader,
  Jenkins, Twicken, Catanzarite, Smith, Batalha, Haas, Thompson, Campbell,
  Sabale, \& Kamal~Uddin}]{Christiansen:2015ge}
Christiansen, J.~L., Clarke, B.~D., Burke, C.~J., {et~al.} 2015, Astrophysical
  Journal, 810, 95

\bibitem[{Ciesla {et~al.}(2015)Ciesla, Mulders, Pascucci, \&
  Apai}]{Ciesla:2015ha}
Ciesla, F.~J., Mulders, G.~D., Pascucci, I., \& Apai, D. 2015, Astrophysical
  Journal, 804, 9

\bibitem[{Cossou {et~al.}(2014)Cossou, Raymond, Hersant, \&
  Pierens}]{2014A&A...569A..56C}
Cossou, C., Raymond, S.~N., Hersant, F., \& Pierens, A. 2014, Astronomy {\&}
  Astrophysics, 569, 56

\bibitem[{Cumming {et~al.}(2008)Cumming, Butler, Marcy, Vogt, Wright, \&
  Fischer}]{Cumming:2008hg}
Cumming, A., Butler, R.~P., Marcy, G.~W., {et~al.} 2008, Publications of the
  Astronomical Society of the Pacific, 120, 531

\bibitem[{Dawson {et~al.}(2015)Dawson, Chiang, \& Lee}]{Dawson:2015ha}
Dawson, R.~I., Chiang, E., \& Lee, E.~J. 2015, Monthly Notices of the Royal
  Astronomical Society, 453, 1471

\bibitem[{Dressing \& Charbonneau(2013)}]{2013ApJ...767...95D}
Dressing, C., \& Charbonneau, D. 2013, The Astrophysical Journal, 767, 95

\bibitem[{Dressing \& Charbonneau(2015)}]{2015ApJ...807...45D}
---. 2015, The Astrophysical Journal, 807, 45

\bibitem[{Dressing {et~al.}(2015)Dressing, Charbonneau, Dumusque, Gettel, Pepe,
  Collier~Cameron, Latham, Molinari, Udry, Affer, Bonomo, Buchhave, Cosentino,
  Figueira, Fiorenzano, Harutyunyan, Haywood, Johnson, Lopez-Morales, Lovis,
  Malavolta, Mayor, Micela, Motalebi, Nascimbeni, Phillips, Piotto, Pollacco,
  Queloz, Rice, Sasselov, Segransan, Sozzetti, Szentgyorgyi, \&
  Watson}]{Dressing:2015je}
Dressing, C., Charbonneau, D., Dumusque, X., {et~al.} 2015, Astrophysical
  Journal, 800, 135

\bibitem[{Fressin {et~al.}(2013)Fressin, Torres, Charbonneau, Bryson,
  Christiansen, Dressing, Jenkins, Walkowicz, \& Batalha}]{2013ApJ...766...81F}
Fressin, F., Torres, G., Charbonneau, D., {et~al.} 2013, The Astrophysical
  Journal, 766, 81

\bibitem[{Goldreich \& Tremaine(1979)}]{Goldreich:1979iq}
Goldreich, P., \& Tremaine, S. 1979, Astrophysical Journal, 233, 857

\bibitem[{Hansen \& Murray(2012)}]{2012ApJ...751..158H}
Hansen, B. M.~S., \& Murray, N. 2012, The Astrophysical Journal, 751, 158

\bibitem[{Hansen \& Murray(2013)}]{2013ApJ...775...53H}
---. 2013, The Astrophysical Journal, 775, 53

\bibitem[{Hasegawa \& Pudritz(2012)}]{2012ApJ...760..117H}
Hasegawa, Y., \& Pudritz, R.~E. 2012, The Astrophysical Journal, 760, 117

\bibitem[{Hasegawa \& Pudritz(2013)}]{2013ApJ...778...78H}
---. 2013, The Astrophysical Journal, 778, 78

\bibitem[{Howard {et~al.}(2010)Howard, Marcy, Johnson, Fischer, Wright,
  Isaacson, Valenti, Anderson, Lin, \& Ida}]{2010Sci...330..653H}
Howard, A.~W., Marcy, G.~W., Johnson, J.~A., {et~al.} 2010, Science, 330, 653

\bibitem[{Howard {et~al.}(2012)Howard, Marcy, Bryson, Jenkins, Rowe, Batalha,
  Borucki, Koch, Dunham, Gautier, Van~Cleve, Cochran, Latham, Lissauer, Torres,
  Brown, Gilliland, Buchhave, Caldwell, Christensen-Dalsgaard, Ciardi, Fressin,
  Haas, Howell, Kjeldsen, Seager, Rogers, Sasselov, Steffen, Basri,
  Charbonneau, Christiansen, Clarke, Dupree, Fabrycky, Fischer, Ford, Fortney,
  Tarter, Girouard, Holman, Johnson, Klaus, Machalek, Moorhead, Morehead,
  Ragozzine, Tenenbaum, Twicken, Quinn, Isaacson, Shporer, Lucas, Walkowicz,
  Welsh, Boss, DeVore, Gould, Smith, Morris, Prsa, Morton, Still, Thompson,
  Mullally, Endl, \& MacQueen}]{2012ApJS..201...15H}
Howard, A.~W., Marcy, G.~W., Bryson, S.~T., {et~al.} 2012, The Astrophysical
  Journal Supplement, 201, 15

\bibitem[{Huber {et~al.}(2014)Huber, Silva~Aguirre, Matthews, Pinsonneault,
  Gaidos, Garc{\'\i}a, Hekker, Mathur, Mosser, Torres, Bastien, Basu, Bedding,
  Chaplin, Demory, Fleming, Guo, Mann, Rowe, Serenelli, Smith, \&
  Stello}]{Huber:2014dh}
Huber, D., Silva~Aguirre, V., Matthews, J.~M., {et~al.} 2014, The Astrophysical
  Journal Supplement Series, 211, 2

\bibitem[{Ida \& Lin(2004)}]{Ida:2004jo}
Ida, S., \& Lin, D. N.~C. 2004, Astrophysical Journal, 616, 567

\bibitem[{Ida \& Lin(2005)}]{Ida:2005bm}
---. 2005, Astrophysical Journal, 626, 1045

\bibitem[{Ida \& Lin(2010)}]{Ida:2010ki}
---. 2010, Astrophysical Journal, 719, 810

\bibitem[{Izidoro {et~al.}(2015)Izidoro, Raymond, Morbidelli, Hersant, \&
  Pierens}]{Izidoro:2015co}
Izidoro, A., Raymond, S.~N., Morbidelli, A., Hersant, F., \& Pierens, A. 2015,
  Astrophysical Journal, 800, L22

\bibitem[{Johnson {et~al.}(2010)Johnson, Aller, Howard, \&
  Crepp}]{Johnson:2010gu}
Johnson, J.~A., Aller, K.~M., Howard, A.~W., \& Crepp, J.~R. 2010, Publications
  of the Astronomical Society of the Pacific, 122, 905

\bibitem[{Kokubo \& Ida(2002)}]{Kokubo:2002dz}
Kokubo, E., \& Ida, S. 2002, Astrophysical Journal, 581, 666

\bibitem[{Kokubo {et~al.}(2006)Kokubo, Kominami, \& Ida}]{2006ApJ...642.1131K}
Kokubo, E., Kominami, J., \& Ida, S. 2006, The Astrophysical Journal, 642, 1131

\bibitem[{Lambrechts \& Johansen(2012)}]{2012A&A...544A..32L}
Lambrechts, M., \& Johansen, A. 2012, Astronomy {\&} Astrophysics, 544, A32

\bibitem[{Laughlin {et~al.}(2004)Laughlin, Bodenheimer, \&
  Adams}]{Laughlin:2004hf}
Laughlin, G., Bodenheimer, P., \& Adams, F.~C. 2004, Astrophysical Journal,
  612, L73

\bibitem[{Lin \& Papaloizou(1979)}]{1979MNRAS.186..799L}
Lin, D. N.~C., \& Papaloizou, J. 1979, Monthly Notices of the Royal
  Astronomical Society, 186, 799

\bibitem[{Lissauer {et~al.}(2011)Lissauer, Ragozzine, Fabrycky, Steffen, Ford,
  Jenkins, Shporer, Holman, Rowe, Quintana, Batalha, Borucki, Bryson, Caldwell,
  Carter, Ciardi, Dunham, Fortney, Gautier, Howell, Koch, Latham, Marcy,
  Morehead, \& Sasselov}]{Lissauer:2011fj}
Lissauer, J.~J., Ragozzine, D., Fabrycky, D.~C., {et~al.} 2011, The
  Astrophysical Journal Supplement Series, 197, 8

\bibitem[{Lopez \& Fortney(2014)}]{Lopez:2014er}
Lopez, E.~D., \& Fortney, J.~J. 2014, Astrophysical Journal, 792, 1

\bibitem[{Luhman \& Mamajek(2012)}]{2012ApJ...758...31L}
Luhman, K.~L., \& Mamajek, E.~E. 2012, The Astrophysical Journal, 758, 31

\bibitem[{Masset {et~al.}(2006)Masset, Morbidelli, Crida, \&
  Ferreira}]{Masset:2006dv}
Masset, F.~S., Morbidelli, A., Crida, A., \& Ferreira, J. 2006, Astrophysical
  Journal, 642, 478

\bibitem[{Miller \& Fortney(2011)}]{Miller:2011bq}
Miller, N., \& Fortney, J.~J. 2011, Astrophysical Journal, 736, L29

\bibitem[{Mohanty {et~al.}(2013)Mohanty, Greaves, Mortlock, Pascucci, Scholz,
  Thompson, Apai, Lodato, \& Looper}]{2013ApJ...773..168M}
Mohanty, S., Greaves, J., Mortlock, D., {et~al.} 2013, The Astrophysical
  Journal, 773, 168

\bibitem[{Montet {et~al.}(2014)Montet, Crepp, Johnson, Howard, \&
  Marcy}]{Montet:2014fa}
Montet, B.~T., Crepp, J.~R., Johnson, J.~A., Howard, A.~W., \& Marcy, G.~W.
  2014, Astrophysical Journal, 781, 28

\bibitem[{Morbidelli {et~al.}(2015)Morbidelli, Lambrechts, Jacobson, \&
  Bitsch}]{2015arXiv150601666M}
Morbidelli, A., Lambrechts, M., Jacobson, S., \& Bitsch, B. 2015, arXiv.org,
  1666

\bibitem[{Morton \& Swift(2014)}]{2014ApJ...791...10M}
Morton, T.~D., \& Swift, J. 2014, The Astrophysical Journal, 791, 10

\bibitem[{Muirhead {et~al.}(2015)Muirhead, Mann, Vanderburg, Morton, Kraus,
  Ireland, Swift, Feiden, Gaidos, \& Gazak}]{Muirhead:2015gw}
Muirhead, P.~S., Mann, A.~W., Vanderburg, A., {et~al.} 2015, Astrophysical
  Journal, 801, 18

\bibitem[{Mulders {et~al.}(2015)Mulders, Pascucci, \& Apai}]{Mulders:2015ja}
Mulders, G.~D., Pascucci, I., \& Apai, D. 2015, Astrophysical Journal, 798, 112

\bibitem[{Mullally {et~al.}(2015)Mullally, Coughlin, Thompson, Rowe, Burke,
  Latham, Batalha, Bryson, Christiansen, Henze, Ofir, Quarles, Shporer,
  Van~Eylen, Van~Laerhoven, Shah, Wolfgang, Chaplin, Xie, Akeson, Argabright,
  Bachtell, Barclay, Borucki, Caldwell, Campbell, Catanzarite, Cochran, Duren,
  Fleming, Fraquelli, Girouard, Haas, He{\l}miniak, Howell, Huber, Larson,
  Gautier, Jenkins, Li, Lissauer, McArthur, Miller, Morris, Patil-Sabale,
  Plavchan, Putnam, Quintana, Ramirez, Silva~Aguirre, Seader, Smith, Steffen,
  Stewart, Stober, Still, Tenenbaum, Troeltzsch, Twicken, \&
  Zamudio}]{Mullally:2015iq}
Mullally, F., Coughlin, J.~L., Thompson, S.~E., {et~al.} 2015, The
  Astrophysical Journal Supplement Series, 217, 31

\bibitem[{Najita \& Kenyon(2014)}]{2014MNRAS.445.3315N}
Najita, J.~R., \& Kenyon, S.~J. 2014, Monthly Notices of the Royal Astronomical
  Society, 445, 3315

\bibitem[{Pascucci {et~al.}(2013)Pascucci, Herczeg, Carr, \&
  Bruderer}]{Pascucci:2013ei}
Pascucci, I., Herczeg, G., Carr, J.~S., \& Bruderer, S. 2013, Astrophysical
  Journal, 779, 178

\bibitem[{Pecaut \& Mamajek(2013)}]{Pecaut:2013ej}
Pecaut, M.~J., \& Mamajek, E.~E. 2013, The Astrophysical Journal Supplement
  Series, 208, 9

\bibitem[{Petigura {et~al.}(2013)Petigura, Marcy, \&
  Howard}]{2013ApJ...770...69P}
Petigura, E.~A., Marcy, G.~W., \& Howard, A.~W. 2013, The Astrophysical
  Journal, 770, 69

\bibitem[{Pinilla {et~al.}(2013)Pinilla, Birnstiel, Benisty, Ricci, Natta,
  Dullemond, Dominik, \& Testi}]{2013A&A...554A..95P}
Pinilla, P., Birnstiel, T., Benisty, M., {et~al.} 2013, Astronomy {\&}
  Astrophysics, 554, A95

\bibitem[{Raghavan {et~al.}(2010)Raghavan, McAlister, Henry, Latham, Marcy,
  Mason, Gies, White, \& ten Brummelaar}]{2010ApJS..190....1R}
Raghavan, D., McAlister, H.~A., Henry, T.~J., {et~al.} 2010, The Astrophysical
  Journal Supplement, 190, 1

\bibitem[{Raymond {et~al.}(2007)Raymond, Scalo, \&
  Meadows}]{2007ApJ...669..606R}
Raymond, S.~N., Scalo, J., \& Meadows, V.~S. 2007, The Astrophysical Journal,
  669, 606

\bibitem[{Rogers(2015)}]{Rogers:2015jn}
Rogers, L.~A. 2015, Astrophysical Journal, 801, 41

\bibitem[{Sanchis-Ojeda {et~al.}(2014)Sanchis-Ojeda, Rappaport, Winn, Kotson,
  Levine, \& El~Mellah}]{SanchisOjeda:2014gi}
Sanchis-Ojeda, R., Rappaport, S., Winn, J.~N., {et~al.} 2014, Astrophysical
  Journal, 787, 47

\bibitem[{Santos {et~al.}(2000)Santos, Israelian, \&
  Mayor}]{2000A&A...363..228S}
Santos, N.~C., Israelian, G., \& Mayor, M. 2000, Astronomy {\&} Astrophysics,
  363, 228

\bibitem[{Schlaufman(2015)}]{Schlaufman:2015di}
Schlaufman, K.~C. 2015, Astrophysical Journal, 799, L26

\bibitem[{Sousa {et~al.}(2008)Sousa, Santos, Mayor, Udry, Casagrande,
  Israelian, Pepe, Queloz, \& Monteiro}]{2008A&A...487..373S}
Sousa, S.~G., Santos, N.~C., Mayor, M., {et~al.} 2008, Astronomy {\&}
  Astrophysics, 487, 373

\bibitem[{Swift {et~al.}(2013)Swift, Johnson, Morton, Crepp, Montet, Fabrycky,
  \& Muirhead}]{2013ApJ...764..105S}
Swift, J.~J., Johnson, J.~A., Morton, T.~D., {et~al.} 2013, The Astrophysical
  Journal, 764, 105

\bibitem[{Tanaka {et~al.}(2002)Tanaka, Takeuchi, \& Ward}]{Tanaka:2002ev}
Tanaka, H., Takeuchi, T., \& Ward, W.~R. 2002, Astrophysical Journal, 565, 1257

\bibitem[{Weidenschilling(1977)}]{1977MNRAS.180...57W}
Weidenschilling, S.~J. 1977, Monthly Notices of the Royal Astronomical Society,
  180, 57

\bibitem[{Weiss \& Marcy(2014)}]{2014ApJ...783L...6W}
Weiss, L.~M., \& Marcy, G.~W. 2014, The Astrophysical Journal Letters, 783, L6

\bibitem[{Wetherill(1996)}]{Wetherill:1996ga}
Wetherill, G.~W. 1996, Astrophysics and Space Science, 241, 25

\bibitem[{Wolfgang \& Lopez(2014)}]{Wolfgang:2014uq}
Wolfgang, A., \& Lopez, E. 2014, eprint arXiv:1409.2982, 1409.2982

\bibitem[{Wolfgang {et~al.}(2015)Wolfgang, Rogers, \&
  Ford}]{2015arXiv150407557W}
Wolfgang, A., Rogers, L.~A., \& Ford, E.~B. 2015, arXiv.org, 7557

\bibitem[{Wu \& Lithwick(2013)}]{Wu:2013cp}
Wu, Y., \& Lithwick, Y. 2013, Astrophysical Journal, 772, 74

\bibitem[{Zeng \& Sasselov(2013)}]{Zeng:2013cs}
Zeng, L., \& Sasselov, D. 2013, Publications of the Astronomical Society of the
  Pacific, 125, 227

\end{thebibliography}
\fi

\end{document}